\documentclass[%
 reprint,
 amsmath,amssymb,
 aps,
]{revtex4-1}

\usepackage{amsmath}
\usepackage{graphicx,epstopdf}
\usepackage{dcolumn}
\usepackage{bm}
\usepackage{xcolor}
\usepackage{float}
\usepackage{enumerate}
\usepackage{natbib}
\usepackage{multirow}

\begin{document}

\title{Phonon hydrodynamics in crystalline GeTe at low temperature}

\author{Kanka Ghosh}
\email{Corresponding author: kanka.ghosh@u-bordeaux.fr}
\affiliation{University of Bordeaux, CNRS, Arts et Metiers Institute of Technology, Bordeaux INP, INRAE, I2M Bordeaux, F-33400 Talence, France}
\author{Andrzej Kusiak}%
\affiliation{University of Bordeaux, CNRS, Arts et Metiers Institute of Technology, Bordeaux INP, INRAE, I2M Bordeaux, F-33400 Talence, France}
 \author{Jean-Luc Battaglia}%
 \affiliation{University of Bordeaux, CNRS, Arts et Metiers Institute of Technology, Bordeaux INP, INRAE, I2M Bordeaux, F-33400 Talence, France}


\begin{abstract}

A first-principles density functional method along with the direct solution of linearized Boltzmann transport equations are employed to systematically analyze the low-temperature thermal transport in crystalline GeTe. The extensive thermal transport simulations, ranging from room temperature to cryogenic temperatures, reveal the emergence of a phonon hydrodynamic regime in GeTe at low temperature. The reduction of grain boundary scattering is found to play a crucial role along with the divergent trend of umklapp and normal scattering at low temperatures in accommodating the hydrodynamic regime. Average scattering rates for normal, umklapp, and other resistive processes are distinguished for a wide range (4–300 K) of temperatures and used for identifying various phonon transport regimes. Therefore, the variations of lattice thermal conductivity, phonon propagation length, and thermal diffusivity with temperature, related to these transport regimes (ballistic, hydrodynamic, and kinetic), have been thoroughly investigated. The mode-wise decomposition of lattice thermal conductivity
and the distinction of thermal diffusivity according to different scattering processes reveal rich information on the dominant phonon modes and phonon scattering processes in GeTe at low temperature. Further, the kinetic-collective model is used to elucidate the hydrodynamic behavior of phonon scattering through the relative study of collective and kinetic contributions to the thermal transport properties. In this context, the Knudsen number is estimated through the characteristic non-local length and the grain size, which further quantifies the consistent hydrodynamic behavior of phonon thermal transport for GeTe. Finally, phonon-vacancy scattering for GeTe is realized, and vacancies are found strongly to influence the hydrodynamic window while incorporating the other resistive scattering mechanisms.
\end{abstract}

\maketitle


\section{Introduction}

Low-temperature phonon-based heat conduction of materials offers some interesting phenomena with its intriguing physics implications, which have drawn sizable attention very recently in the field of phonon heat transport \cite{Lucas2019, Bi2018, Cepellotti2015, Torres_2019, Koreeda2010}. One such phenomenon is phonon hydrodynamics, which deals with the collective motion of phonons as a medium of heat conduction and bears conceptual similarities with the hydrodynamic fluid flow, contrary to the usual single-mode relaxation-time approximation, where the energy and lifetime of each independent phonon gas particles are considered \cite{Leebookchapter1, Lucas2019, Cepellotti2015, Hardy, GUO20151, Machidaeaat3374}. The collective flow of phonons is caused by the strong presence of normal scattering ($N$) events, which allow the phonons to conserve their momentum before being dissipated by weak resistive ($R$ = umklapp, phonon-boundary, or phonon-isotope scattering) scattering events \cite{Leebookchapter1, Lucas2019,Bi2018, Cepellotti2015, GUO20151}. As a result, under certain conditions, phonons manifest characteristic length and time scales over which temperature fluctuations propagate as damped waves \cite{Bi2018, Transport_waves_as_crystal_excitations} and feature exotic phenomena like Poiseuille’s flow and the occurrence of second sound\cite{umklapp_gang_2018, Ding2018, Cepellotti2015, Machidaeaat3374, Guyer1966_2}. The idea of identifying the phonon hydrodynamic regime using the average scattering rate of normal and other momentum-destroying resistive scattering processes, was first proposed by Guyer and Krumhansl through their seminal theoretical work \cite{Guyer1966_1, Guyer1966_2}. From a different perspective, hydrodynamic effects have also been understood from their deviation from Fourier’s law behavior of phonon thermal transport \cite{Giorgia2014nanoletters, Gill2015}. Recently, phonon collective excitations have been treated differently by defining them as relaxons, an elementary carrier of heat that is defined as the eigenvectors of the scattering matrix \cite{Relaxon}.

Until now, only a handful of materials, mostly two-dimensional (2D) materials, have been found to exhibit phonon hydrodynamics both theoretically and experimentally \cite{Lucas2019, Gill2015, Li_graphene2018, Li_graphene_2019, Derek_graphene_2018}. First-principles simulations by A. Cepellotti and co-workers \cite{Cepellotti2015} suggested that the hydrodynamic effects can be observed even at room temperature for graphene, boron nitride and other 2D materials. Further, the existence of second sound had also been realized through lattice dynamics calculations for a single-walled carbon nanotube \cite{Lee2017CNT}. Recently S. Huberman \textit{et al}. \cite{Huberman375} experimentally observed second sound in graphite above 100 K, validating the earlier theoretical first-principles study by Z. Ding \textit{et al.} \cite{Ding2018}. Very recently, the relation between the thickness and thermal conductivity, and therefore the link between these two factors with phonon hydrodynamics was studied for graphite \cite{Machida309}. A faster than T$^3$ scaling of the lattice thermal conductivity has also been identified as an observation to detect phonon hydrodynamics in recent studies comprised of both experimental and theoretical methods on bulk black phosphorus \cite{Machidaeaat3374} and SrTiO$_{3}$ \cite{Koreeda2007, Strontium2018}. Low frequency light-scattering and time-domain light-scattering techniques were also employed to study collective phonon excitation in KTaO$_3$, and the corresponding hydrodynamic behavior was observed below 30 K \cite{Koreeda2010}. The presence of second sound was experimentally observed at low temperatures in isotopically pure solid helium (0.6-1 K) \cite{Solid_Helium}, NaF ($\sim$ 15 K) \cite{NaF} and Bi ($\sim$ 3 K) \cite{Bi1972}. Recently, theoretical calculations performed by M. Markov \textit{et al}. \cite{Bi2018} confirmed the experimental realization of hydrodynamic Poiseuille phonon flow in bismuth (Bi) at cryogenic temperature.

Germanium telluride (GeTe) is a versatile material with its  diverse range of applicability \cite{Levin2013, Campi2015}. Due to its notably high contrast in electrical resistance and a stable amorphous phase with a higher crystallization temperature, it has emerged as one of the most significant candidates within phase change materials \cite{Andrzej,Urszula2014}. GeTe has been implemented with a superlattice configuration as GeTe-Sb$_{2}$Te$_{3}$ , which has been broadly used for its application in optical as well as PCM storage devices \cite{Boschker, Campi2015}. It is also an efficient thermoelectric material and is used for applications in waste heat recovery, low-scale refrigeration, and energy generation etc \cite{Levin2013}. Most of the works on the thermal conductivity of crystalline GeTe has been carried out at room temperature \cite{Levin2013, Campi2015, Campi2017, Ronald2019}, at high temperatures \cite{Levin2013} and within a range from room temperature to high temperature \cite{Nath, kanka} mostly due to its engineering applications. However, at low temperature, very few investigations have been done to understand the heat transfer mechanism. Several decades ago, Lewis \textit{et al.} \cite{Lewis1968} experimentally measured the thermal conductivity of GeTe in the temperature interval of 2.5-110 K. Recently, the lattice thermal conductivity of arc-melted Ge-deficient GeTe was experimentally measured \cite{Sanchez2018} in the temperature range of 10-800K. However, no physical insight was provided to understand the illusive role of phonon scattering at low temperatures. In a recent study, Torres \textit{et al.} \cite{Torres_2019} showed a strong phonon hydrodynamic behavior in low lattice thermal conductivity ($\kappa_L$) materials such as metal dichalcogenides. Therefore, we ask the following question: Can GeTe, a chalcogenide-based material, which shows even lower lattice thermal conductivity ($\kappa_L$) compared to metal dichalcogenides, exhibit appreciable phonon hydrodynamics at low temperatures ? Also, featuring a considerable hydrodynamic effect in a material demands simultaneous weak and strong umklapp and normal scattering respectively at low temperature. However, inadequate normal scattering events influence the phonon transport to become ballistic \cite{Leebookchapter1}. Having understood the distinct role of different scattering mechanisms as well as various phonon modes in the thermal transport of GeTe at temperatures ranging from room temperature to 503 K in our recent study \cite{kanka}, we tend to understand the hierarchy of phonon scattering mechanisms and their implications at low temperature.

Therefore, in this current paper, we explore the low temperature thermal transport of crystalline GeTe, ranging from 4 to 300 K, using the first-principles density functional method coupled with the solution of the linearized Boltzmann transport equation (LBTE) via a direct non-iterative method. To compare and investigate the regime of failure of a phonon gas model and an individual phonon scattering description at low temperature, the relaxation-time approximation (RTA) is also studied. After defining average scattering rates in the investigated temperature range, a systematic study of mode-decomposed lattice thermal conductivity is carried out. Two different grain-sizes have been considered to understand the role of phonon-boundary scattering. Ballistic, hydrodynamic and kinetic transport regimes are identified. The variation of second sound propagation length with temperature has been discussed and compared with the phonon average mean free path. Thermal diffusivity and its contribution from different scattering events have been estimated. To get further insight and consistency, the kinetic-collective model (KCM) is employed and the relative contribution of collective and kinetic thermal transport has been understood from lattice thermal conductivity and Knudsen number estimation. Finally, phonon-vacancy scattering for GeTe is studied which was found to affect the hydrodynamic regime of GeTe. This thorough and systematic in-depth theoretical investigations and its findings are crucial to understand the illusive nature and hierarchy of different phonon scattering events  for chalcogenide low-$\kappa_L$ materials.

\section{Computational Details}{\label{section:computational}}
The structural parameters of crystalline GeTe (space group $R3m$) are optimized via first-principles density functional calculations, and the corresponding parameter details are presented in our earlier work \cite{kanka}. The phonon lifetime and consequently the lattice thermal conductivity $\kappa_L$ are obtained by solving linearized phonon Boltzmann transport equation (LBTE), using both direct method introduced by L. Chaput et al. \cite{Chaput} as well as the single mode relaxation time approximation (RTA) or the RTA, employing PHONO3PY \cite{Togo} software package. Initially, the supercell approach with finite displacement of 0.03 \AA{} is applied to calculate the harmonic (second order) and the anharmonic (third order) force constants, given by 
\begin{equation}
    \Phi_{\alpha \beta} (l\kappa, l'\kappa') = \frac{\partial^2 \Phi}{\partial u_{\alpha} (l\kappa)\partial u_{\beta} (l'\kappa')}
\end{equation}
and
\begin{equation}
    \Phi_{\alpha \beta \gamma} (l\kappa, l'\kappa', l''\kappa'') = \frac{\partial^3 \Phi}{\partial u_{\alpha} (l\kappa)\partial u_{\beta} (l'\kappa') \partial u_{\gamma} (l''\kappa'')}
\end{equation}
respectively. First principles calculations using QUANTUM-ESPRESSO \cite{qe} are implemented to calculate the forces acting on atoms in supercells.  Using finite difference method, harmonic force constants are approximated as \cite{Togo} 
\begin{equation}
   \Phi_{\alpha \beta} (l\kappa, l'\kappa') \simeq - \frac{F_{\beta} [l'\kappa'; \textbf{u} (l \kappa)]}{u_{\alpha} (l \kappa)} 
\end{equation}
where \textbf{F}[$l'$$\kappa'$; \textbf{u}($l$$\kappa$)] is atomic force computed at \textbf{r}($l'$ $\kappa'$) with an atomic displacement \textbf{u}($l\kappa$) in a supercell. Similarly, anharmonic force constants are obtained using\cite{Togo} 
\begin{equation}
   \Phi_{\alpha \beta \gamma} (l\kappa, l'\kappa', l''\kappa'') \simeq - \frac{F_{\gamma} [l''\kappa''; \textbf{u} (l \kappa), \textbf{u} (l' \kappa')]}{u_{\alpha} (l \kappa)u_{\beta} (l' \kappa')} 
\end{equation}
where \textbf{F}[$l''$$\kappa''$; \textbf{u}($l$$\kappa$), \textbf{u}($l'$ $\kappa'$)] is the atomic force computed at \textbf{r}($l''$ $\kappa''$) with a pair of atomic displacements \textbf{u}($l\kappa$) and \textbf{u}($l'\kappa'$) in a supercell. These two sets of linear equations are solved using the Moore-Penrose pseudoinverse as is implemented in PHONO3PY \cite{Togo}.

We use a 2$\times$2$\times$2 supercell of GeTe for our first-principles calculations of anharmonic force constants. Using the supercell and finite displacement approach, 228 supercells are obtained, having different pairs of displaced atoms, for the calculations for the anharmonic force constants. A larger 3$\times$3$\times$3 supercell is employed for calculating the harmonic force constants. For all the supercell force calculations, the reciprocal space is sampled using a 3$\times$3$\times$3 k-sampling Monkhorst-Pack (MP) mesh \cite{MP} shifted by a half-grid distances along all three directions from the $\Gamma$- point. For the density functional calculations, the Perdew-Burke-Ernzerhof (PBE) \cite{PBE} generalized gradient approximation (GGA) is used as the exchange-correlation functional. The spin-orbit interaction has been ignored due to its negligible effects on the vibrational features of GeTe as mentioned in the literature \cite{Shaltaf2009,Campi2017}. Electron-ion interactions are represented by pseudopotentials using the framework of the projector-augmented-wave (PAW) method \cite{PAW}. The Kohn-Sham (KS) orbitals are expanded in a plane-wave (PW) basis with a kinetic cutoff of 60 Ry and a charge density cutoff of 240 Ry as prescribed by the pseudopotentials of Ge and Te. The total energy convergence threshold has been kept at 10$^{-10}$ a.u. for supercell calculations. For lattice thermal conductivity calculations employing both the direct solution of LBTE and that of the RTA, a $\textbf{q}$-mesh of 24$\times$24$\times$24 is used. The imaginary part of the self-energy has been calculated using the tetrahedron method from which phonon lifetimes are obtained. For KCM \cite{KCM-method2017} calculations, KCM.PY code \cite{KCM-method2017} with the PHONO3PY \cite{Togo} implementation is employed.

\section{Average phonon scattering rate and hydrodynamic regime}{\label{section:average_phonon_scattering}}

The theory of lattice dynamics assumes crystal potential energy to be an analytical function of the atomic displacements from their equilibrium positions \cite{Togo}. Therefore, the crystal potential is expanded with respect to atomic displacements and the corresponding third order coefficients that contain the anharmonicity are employed to calculate the imaginary part of the self energy \cite{Togo}. The phonon lifetime ($\tau_{ph-ph}$) is computed from the imaginary part of the phonon self energy using PHONO3PY \cite{Togo, Mizokami}.
Generally, in a harmonic approximation, phonon lifetimes are infinite whereas, anharmonicity in a crystal gives rise to a phonon self energy  $\Delta\omega_{\lambda}$ + $i\Gamma_{\lambda}$. The phonon lifetime has been computed from the imaginary part of the phonon self energy as $\tau_{\lambda}$ = $\frac{1}{2\Gamma_{\lambda}(\omega_{\lambda})}$ from\cite{Togo}
\begin{widetext}
\begin{equation}
    \Gamma_{\lambda}(\omega_{\lambda}) = \frac{18\pi}{\hbar^2}\sum_{\lambda'\lambda''}\Delta\left( \textbf{q}+\textbf{q}'+\textbf{q}'' \right)\mid \Phi_{-\lambda\lambda'\lambda''}\mid ^{2}\{(n_{\lambda'}+n_{\lambda''}+1)\delta(\omega-\omega_{\lambda'}-\omega_{\lambda''}) + (n_{\lambda'}-n_{\lambda''})[\delta(\omega+\omega_{\lambda'}-\omega_{\lambda''}) - \delta(\omega-\omega_{\lambda'}+\omega_{\lambda''})]\}
\end{equation}
\end{widetext}
where $n_\lambda$ = $\frac{1}{exp(\hbar\omega_{\lambda}/k_{B}T)-1}$ is the phonon occupation number at the equilibrium. $\Delta\left(\textbf{q}+\textbf{q}'+\textbf{q}'' \right)$ = 1 if $\textbf{q}+\textbf{q}'+\textbf{q}'' = \textbf{G}$, or 0 otherwise. Here \textbf{G} represents reciprocal lattice vector. Integration over \textbf{q}-point triplets for the calculation is made separately for normal (\textbf{G} = 0) and umklapp processes (\textbf{G} $\neq$ 0) and therefore phonon umklapp ($\tau_U$) and phonon normal lifetime ($\tau_N$) have been distinguished. 

For both the direct method and RTA, scattering of phonon modes by randomly distributed isotopes \cite{Togo} is also incorporated for comparison. The isotope scattering rate ($\tau_{I}^{-1}$), using second-order perturbation theory, is given by Shin-ichiro Tamura \cite{Tamura} as
\begin{equation}
\resizebox{1.0\hsize}{!}{$
    \frac{1}{\tau_{\lambda}^{I}(\omega)} = \frac{\pi \omega_{\lambda}^{2}}{2N}\sum_{\lambda'} \delta\left(\omega - \omega'_{\lambda} \right) \sum_{k} g_{k}|\sum_{\alpha}\textbf{W}_{\alpha}\left(k,\lambda \right)\textbf{W}_{\alpha}^{*}\left(k,\lambda \right)| ^{2} 
$}
\end{equation}
where $g_k$ is the mass variance parameter, defined as 
\begin{equation}
    g_{k} = \sum_{i} f_{i} \left( 1 - \frac{m_{ik}}{\overline{m}_k}\right)^{2}
\end{equation}
$f_i$ is the mole fraction, $m_{ik}$ is the relative atomic mass of $i$th isotope, $\overline{m}_k$ is the average mass = $\sum_{i} f_{i} m_{ik}$, and $\textbf{W}$ is a polarization vector. The database of the natural abundance data for elements \cite{Laeter} is used for the mass variance parameters.

The effect of a crystal boundary has been implemented using Casimir diffuse boundary scattering \cite{kaviany_2014} as $\tau_{\lambda}^{B}$ = $\frac{L}{\mid \textbf{v}_{\lambda} \mid}$ where, $\textbf{v}_{\lambda}$ is the average phonon group velocity of phonon mode $\lambda$ and $L$ is the grain size, which is also called Casimir length L, the length phonons travel before the boundary absorption or re-emission \cite{kaviany_2014}. 

The thermodynamic average of different phonon scattering events is defined using:
\begin{equation}
    \langle \tau_{i}^{-1} \rangle_{ave} = \frac{\sum_{\lambda}C_{\lambda}\tau_{\lambda i}^{-1}}{\sum_{\lambda}C_{\lambda}}
\end{equation}
Here, $\lambda$ defines phonon modes ($\textbf{q}$, $j$) comprising wave vector $\textbf{q}$ and branch $j$. Index $i$ denotes normal, umklapp, isotope and boundary scattering processes used, denoted by N, U and I and B respectively. $C_\lambda$ is the modal heat 
\onecolumngrid
\begin{widetext}
\begin{figure}[H]
    \centering
    \includegraphics[width=1.0\textwidth]{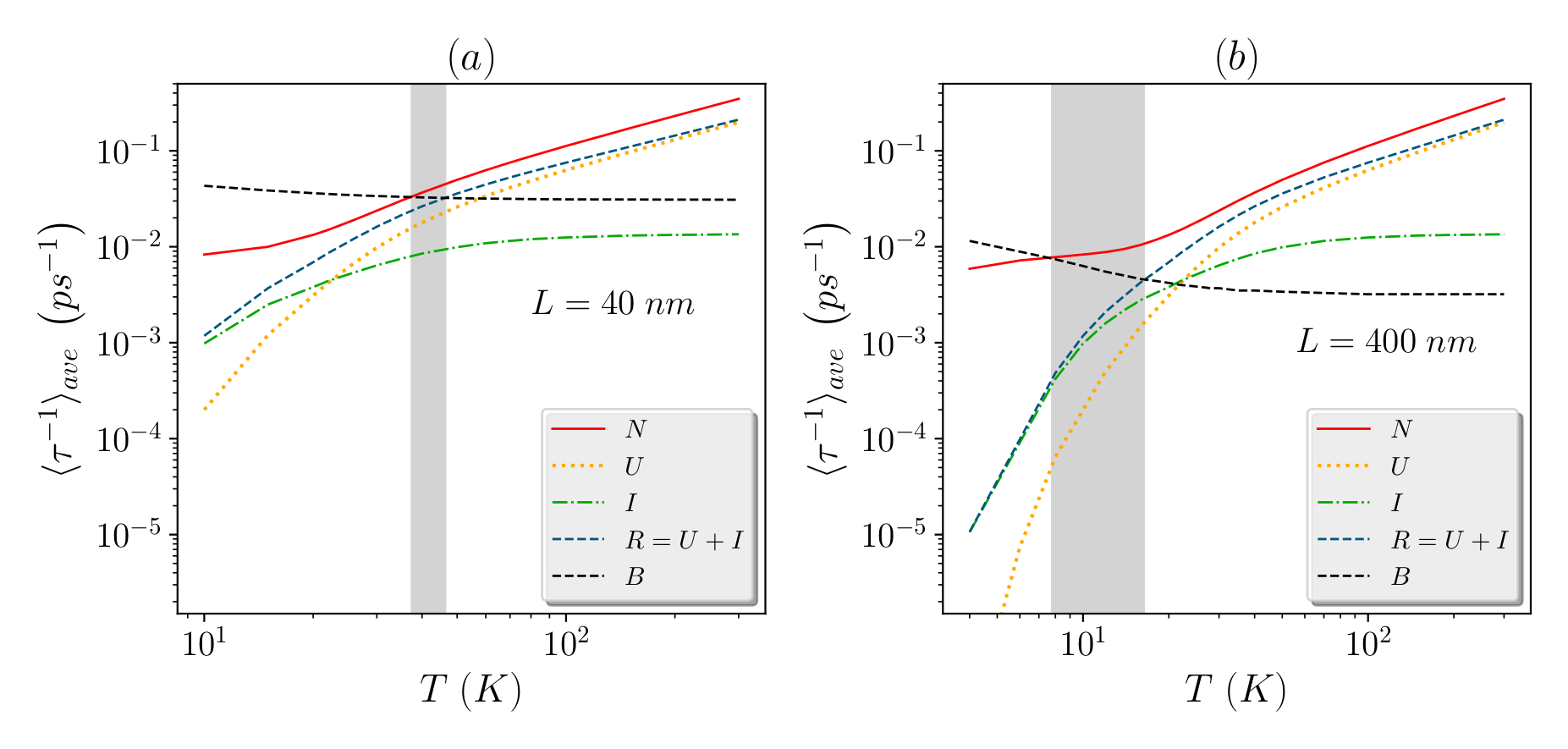}
    \caption{Thermodynamic average phonon scattering rates as a function of temperature in log-log scale for GeTe for (a) grain size ($L$) = 40 nm and (b) grain size ($L$) = 400 nm. $N$, $U$, $I$, $B$ and $R$ denote normal, umklapp, isotope, boundary and resistive scattering respectively. The shaded regions correspond to the validation of the Guyer's condition \cite{Guyer1966_2} for Poiseuille's flow (Eq. \ref{eq:poiseuille})}
    \label{fig:average_phonon_scattering}
\end{figure}
\vspace{-0.3cm}
\end{widetext}
capacity, given by 
\begin{equation}
    C_\lambda = k_{B} \left(\frac{\hbar\omega_{\lambda}}{k_{B}T}\right)^{2} \frac{exp(\hbar\omega_{\lambda}/k_{B}T)}{[exp(\hbar\omega_{\lambda}/k_{B}T) -1]^2}
\end{equation}
where, $T$ denotes temperature, $\hbar$ is the reduced Planck constant and $k_B$ is the Boltzmann constant. 
According to the condition prescribed by Guyer and Krumhansl \cite{Guyer1966_2} hydrodynamic regime exists if
\begin{equation}{\label{eq:hydro}}
   \langle \tau_{U}^{-1} \rangle_{ave} \ll \langle \tau_{N}^{-1} \rangle_{ave}
\end{equation}
Moreover, Guyer's condition \cite{Guyer1966_2} for the occurrence of second sound and Poiseuille's flow reads: 
\begin{equation}{\label{eq:poiseuille}}
   \langle \tau_{U}^{-1} \rangle_{ave} < \langle \tau_{B}^{-1} \rangle_{ave} < \langle\tau_{N}^{-1} \rangle_{ave}
\end{equation}

We carefully introspect these conditions in GeTe for the two different grain sizes. Here we note that our earlier experimental study \cite{Andrzej} on the variation of GeTe grain size with annealing temperature revealed the grain-size to be 40 nm for an annealing temperature of 453 K and the grain size grows with increasing annealing temperature. Therefore, in this investigation, we study two different grain sizes namely 40 nm and 400 nm to study the effect of grain size on the phonon hydrodynamics. 

Figure \ref{fig:average_phonon_scattering} presents the average scattering rates due to various scattering processes as a function of temperature for GeTe. The shaded region in the temperature range defines the regime where Guyer's condition \cite{Guyer1966_2} for Poiseuille's flow (Eq. \ref{eq:poiseuille}) is satisfied. It is observed that at high temperatures, the difference between umklapp and normal scattering is low and it gradually increases with decreasing temperature. The average isotope scattering shows an almost constant value at higher temperature and subsequently a gradual dip in the value as the temperature is lowered. The resistive scattering rate ($\tau_{R}^{-1}$ = $\tau_{U}^{-1}$ + $\tau_{I}^{-1}$) is realized by adding the resistive scattering processes, namely umklapp and phonon-isotope scattering. Similar to the trend of the umklapp scattering rate with temperature, the difference between resistive and normal scattering rate is low at high temperatures and it gradually increases upon lowering the temperature. Here we mention that the average resistive scattering is dominated by umklapp scattering at higher temperatures. However, as temperature is lowered, isotope scattering emerges as a significant contributor for total resistive scattering in GeTe and even dominates the resistive scattering rate at further lowering of temperature. Therefore, as mentioned earlier \cite{Bi2018}, isotopic purity is an important factor for the existence of the hydrodynamic regime and isotopic impurity can reduce the chances of hydrodynamic phonon flow.

Finally, satisfying or not satisfying both of Guyer's conditions (Eq. \ref{eq:hydro} and Eq. \ref{eq:poiseuille}) crucially depends on the phonon-boundary scattering, or more elaborately, the grain size. Equation \ref{eq:poiseuille} has been found to be valid in the temperature regime $\approx$ 37-47 K (Fig \ref{fig:average_phonon_scattering}.(a)) for $L$ = 40 nm, whereas a temperature regime of $\approx$ 8-16 K (Fig \ref{fig:average_phonon_scattering}.(b)) is identified for the Poiseuille flow regime for $L$ = 400 nm. Although for both of the grain sizes, Eq. \ref{eq:poiseuille} is satisfied in the regime defined above, Eq. \ref{eq:hydro} is found to be valid only for $L$ = 400 nm (Fig \ref{fig:average_phonon_scattering}.(b)). Therefore, reducing the phonon-boundary scattering using a larger grain size is found to be an avenue to explore the hydrodynamic regime in GeTe.

\section{Lattice thermal conductivity for different transport regimes: Acoustic and optical mode decomposition}{\label{section:Lattice thermal conductivity}}

After defining the hydrodynamic regime from phonon scattering rates, we tend to investigate the lattice thermal conductivity ($\kappa_L$) of crystalline GeTe as a function of temperature. We note that the lattice thermal conductivity picture can also serve as a way to distinguish different phonon transport regimes when direct non-iterative solutions to Boltzmann transport equations (LBTE) are compared with that of the single mode relaxation time (RTA) solution. The deviation of RTA $\kappa_L$ from the direct solution of LBTE $\kappa_L$ can be understood as a marker to the failure of the concept of a single, uncorrelated phonon heat transfer mechanism \cite{Bi2018, Lucas2019}. 

In order to evaluate the lattice thermal conductivity ($\kappa_L$) through the direct solution of LBTE, the method developed by L. Chaput \cite{Chaput} is adopted. According to this method, lattice thermal conductivity is given as \cite{Chaput} 
\begin{equation}
    \kappa_{\alpha\beta} = \frac{\hbar^2}{4k_{B}T^{2}NV_{0}} \sum_{\lambda\lambda'} \frac{\omega_{\lambda}\upsilon_{\alpha}(\lambda)}{sinh(\frac{\hbar\omega_{\lambda}}{2k_{B}T})}\frac{\omega_{\lambda'}\upsilon_{\beta}(\lambda')}{sinh(\frac{\hbar\omega_{\lambda'}}{2k_{B}T})} (\Omega^{\sim1})_{\lambda\lambda'}
\end{equation}
where, $\Omega^{\sim1}$ is the Moore-Penrose inverse of the collision matrix $\Omega$, given by \cite{Chaput,Togo}
\begin{widetext}
\begin{equation}
    \Omega_{\lambda\lambda'} = \delta_{\lambda\lambda'}/\tau_{\lambda} + \pi/\hbar^{2} \sum_{\lambda''}\mid \Phi_{\lambda\lambda'\lambda''}\mid ^{2}\frac{[\delta(\omega_{\lambda}-\omega_{\lambda'}-\omega_{\lambda''}) + \delta(\omega_{\lambda}+\omega_{\lambda'}-\omega_{\lambda''}) + \delta(\omega_{\lambda}-\omega_{\lambda'}+\omega_{\lambda''})]}{sinh(\frac{\hbar\omega_{\lambda''}}{2k_{B}T})}
\end{equation}
\end{widetext}

Here, $\Phi_{\lambda\lambda'\lambda''}$ denotes the interaction strength between three phonon $\lambda$, $\lambda'$ and $\lambda''$ scattering \cite{Togo}. However, implementing the RTA in solving LBTE, lattice thermal conductivity tensor $\boldsymbol{\kappa}_L$ can be written in a closed form as \cite{Giorgia,Togo}
\begin{equation}{\label{equation_kl}}
    \boldsymbol{\kappa}_L = \frac{1}{NV_0} \sum_{\lambda} C_{\lambda} \textbf{v}_{\lambda} \otimes \textbf{v}_{\lambda}\tau_{\lambda}
\end{equation}
where $N$ is the number of unit cells and $V_0$ is the volume of the unit cell. $C_\lambda$ is the modal heat capacity, $\lambda$ being the mode. We consider different scattering processes, namely normal, umklapp, isotope and boundary scattering denoted by $N$, $U$, $I$ and $B$ respectively. For each of these processes, the total phonon lifetime has been realized using Matthiessen's rule as \cite{kaviany_2014} 
\begin{equation}{\label{matthiessen}}
    \frac{1}{\tau_{\lambda}} = \frac{1}{\tau_{\lambda}^N} + \frac{1}{\tau_{\lambda}^U} + \frac{1}{\tau_{\lambda}^I} + \frac{1}{\tau_{\lambda}^B}
\end{equation}
where $\tau_{\lambda}^N$, $\tau_{\lambda}^U$, $\tau_{\lambda}^I$ and $\tau_{\lambda}^B$ are phonon lifetimes corresponding to the normal, umklapp, isotope and boundary scattering respectively.

Figure \ref{fig:LTC_mode_decomposed}. (a) and (b) present $\kappa_L$ as a function of temperature for grain sizes 40 nm and 400 nm respectively. Since GeTe is polycrystalline and $\kappa_L$ is anisotropic along the hexagonal c axis and a-b axes \cite{Campi2017, kanka}, the average lattice thermal conductivity is calculated as $\kappa_{av}$ = $\frac{2}{3}$$\kappa_x$ + $\frac{1}{3}$$\kappa_z$ \cite{kanka, Campi2017}. We will focus here on the direct solution of LBTE (red circles) and will use RTA solutions for comparison. At higher temperature, $\kappa_L$ is found to follow a $\frac{1}{T}$ trend, reminiscing of the significant contribution from umklapp scattering described by the phenomenological Slack model \cite{Bosoni, slack1964, Morelli2006} and thus defining the kinetic regime (light red shaded region in Fig \ref{fig:LTC_mode_decomposed}) of thermal transport for GeTe. We observe (green dotted lines in Fig \ref{fig:LTC_mode_decomposed}.(a) and (b)) that the extent of the kinetic regime for GeTe is longer for $L$ = 400 nm ($\approx$ 150-300 K) than $L$ = 40 nm ($\approx$ 200-300 K). Lowering the temperature gradually enhances the $\kappa_L$ to reach maximum and then gradually helps dropping the $\kappa_L$ to zero upon further temperature lowering. While reaching maximum is a manifestation of enhanced normal scattering, phonon boundary scattering is responsible for the decrement of $\kappa_L$ from maximum to zero. Thus, a higher value of $\kappa_L$ is observed (Fig \ref{fig:LTC_mode_decomposed}.(b)) for the relatively weak phonon boundary scattering rate by increasing the grain size. 

By carefully comparing the $\kappa_L$-LBTE solution with that of the RTA in Fig \ref{fig:LTC_mode_decomposed}.(a) and (b), a qualitative estimate can be made regarding the hydrodynamic regime. The difference between the direct LBTE solution (red circles) and the RTA (blue line) gradually increases as we lower the temperature from 300 K. After the difference reaches maximum, the two solutions start to fall into each other and become identical at very low temperatures. According to the conventional wisdom \cite{Bi2018,Lucas2019}, the difference becomes maximum when the single uncorrelated phonon gas concept with an individual phonon lifetime breaks down, and momentum conserving normal scattering overpowers resistive scattering processes. Thus deviating from $1/T$ scaling marks the departure from the kinetic phonon transport regime and gradual entrance to the hydrodynamic regime. On the other  
\onecolumngrid
\begin{widetext}
\begin{figure}[H]
    \centering
    \includegraphics[width=1.0\textwidth]{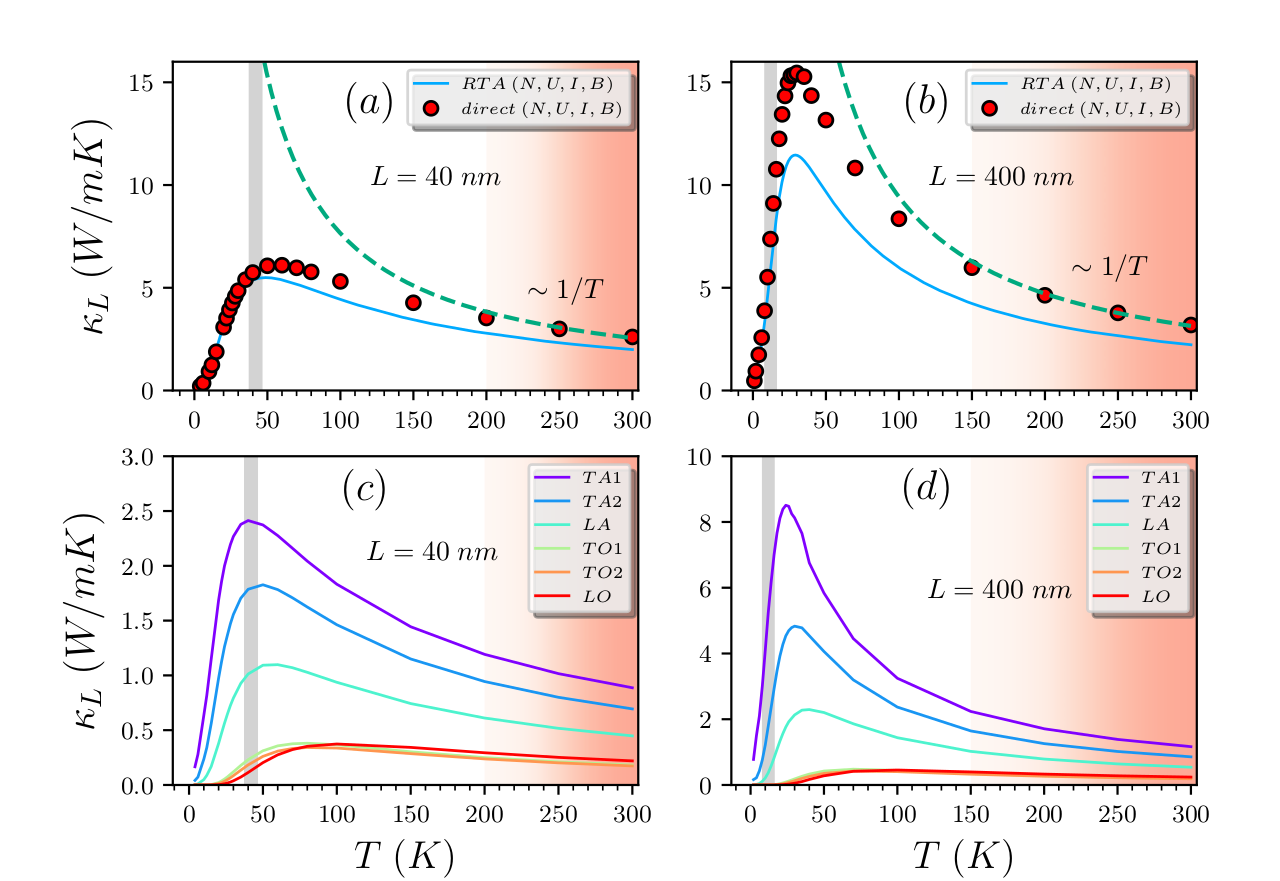}
    \caption{Lattice thermal conductivity ($\kappa_L$) as a function of temperature using direct-LBTE as well as RTA solution for GeTe with grain size (a) 40 nm and (b) 400 nm. Separate contributions of each phonon mode to $\kappa_L$ is presented as a function of temperature using direct-LBTE for grain size (c) 40 nm and (d) 400 nm. Transverse acoustic and optical modes are denoted by TA and TO respectively whereas longitudinal acoustic and optical modes are denoted by LA and LO respectively. The shaded region with light red and gray denote kinetic and hydrodynamic regimes respectively.}
    \label{fig:LTC_mode_decomposed}
\end{figure}
\vspace{-0.3cm}
\end{widetext}
extreme, as shown in Fig \ref{fig:LTC_mode_decomposed}.(a) and (b), at very low temperatures, boundary scattering dominates over any other scattering mechanisms, and direct-LBTE and RTA based solutions become identical. This is called the ``ballistic regime" where the phonon mean free path is dictated by the grain boundary Casimir length $L$. As the temperature is increased, the two solutions start diverging and thus phonon transport enters the hydrodynamic regime from the ballistic regime. We find that the criteria for the hydrodynamic regime and phonon Poiseuille flow, shown as the shaded area in Fig \ref{fig:LTC_mode_decomposed}.(a) and (b), marks consistently the regions for both the grain sizes that are close to the  $\kappa_L$-maximum and start from the point where direct-LBTE and RTA solutions just start diverging immediately after the ballistic regime.

Fig \ref{fig:LTC_mode_decomposed}.(c) and (d) show the variation of each decomposed acoustic and optical modes of $\kappa_L$ with temperature for L = 40 nm and 400 nm respectively. Consistent with the recent observation from the high temperature study of GeTe \cite{kanka}, transverse acoustic modes are found to dominate the phonon heat transfer throughout the whole temperature range studied. The contribution of optical modes happens to be substantially low for GeTe. For $T$ $<$ 30 K, this contribution nearly vanishes (as will be discussed further below).

\begin{widetext}
\begin{figure}[H]
    \centering
    \includegraphics[width=1.0\textwidth]{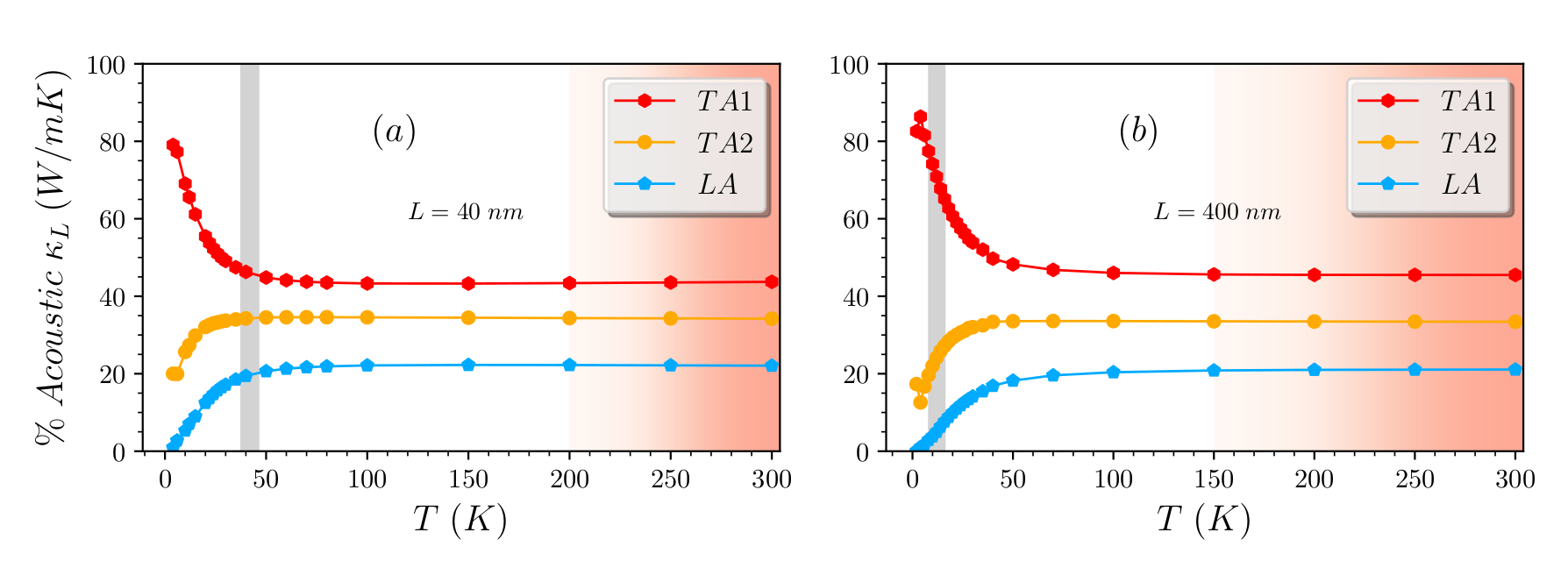}
    \caption{Contribution of transverse (TA1, TA2) and longitudinal (LA) acoustic modes to total $\kappa_L$ for GeTe as a function of temperature for (a) L = 40 nm and (b) L = 400 nm.}
    \label{fig:AM_contrib}
\end{figure}
\end{widetext}

The contributions of three acoustic modes (TA1, TA2 and LA) are found to evolve in a different fashion with temperature as realised via Fig \ref{fig:AM_contrib}. The TA1 mode seems to be the dominant contributor for the whole temperature range for both $L$ = 40 nm and 400 nm. At higher temperature, the contribution of TA1 reaches a constant value of $\approx$ 45 $\%$, while TA2 and LA modes contribute $\approx$ 34 $\%$ and $\approx$ 21 $\%$ respectively to the total acoustic $\kappa_L$. As the temperature is lowered below 100 K, both TA2 and LA contributions start decreasing. Remarkably, the TA1 contribution shows a gradual increasing trend below 100 K and reaches a maximum value around 80 $\%$ at extreme low temperature for both grain sizes. As the hydrodynamic regime is prominent for $L$ = 400 nm, looking   

\onecolumngrid

\vspace{-1cm}
\onecolumngrid
\begin{widetext}
\vspace{-0.5cm}
\begin{figure}[H]
    \centering
    \includegraphics[width=0.75\textwidth]{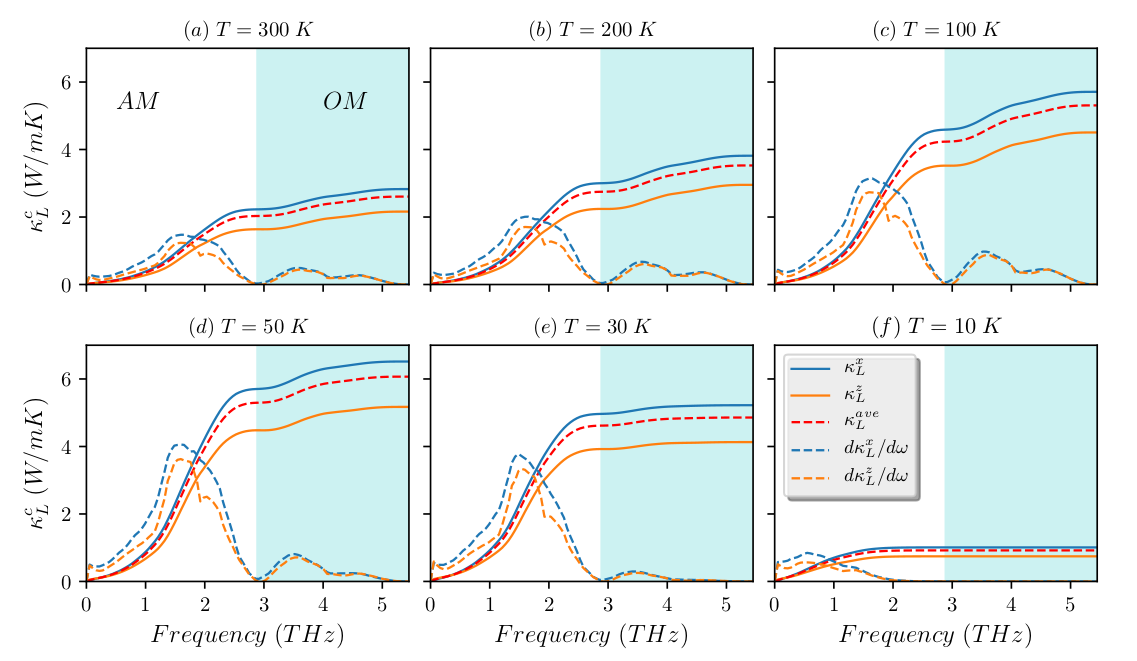}
    \caption{Cumulative lattice thermal conductivities ($\kappa_L^c$) of crystalline GeTe are presented as a function of frequencies at six different temperatures: (a) T = 300 K, (b) T = 200 K, (c) T = 100 K, (d) T = 50 K, (e) T = 30 K and (f) T = 10 K, for L = 40 nm. AM and OM define acoustic and optical modes respectively. Cumulative $\kappa_L^c$, computed along hexagonal c axis ($\kappa_L^z$), along its perpendicular direction ($\kappa_L^x$) and their average $\kappa_L^{av}$ are shown. The derivatives of $\kappa_L^z$ and $\kappa_L^x$ with respect to frequencies are also shown for each temperature.}
    \label{fig:cumulativeAMOM}
\end{figure}
\end{widetext}
at Fig \ref{fig:AM_contrib}.(b), we can observe that the contributions of all 3 acoustic modes are constant in the kinetic regime, whereas, in the hydrodynamic and ballistic transport regimes, the TA1 mode overshadows the TA2 and LA modes. To further understand the role played by the optical modes, compared to the acoustic modes, in thermal transport in GeTe, we calculate the cumulative lattice thermal conductivity ($\kappa_{L}^{c}$) as a function of phonon frequency defined as \cite{Togo,Mizokami}
\begin{equation}
    \kappa_L^c = \int_0^{\omega} \kappa_{L} (\omega')d\omega'
\end{equation}
\begin{figure}[H]
    \centering
    \includegraphics[width=0.5\textwidth]{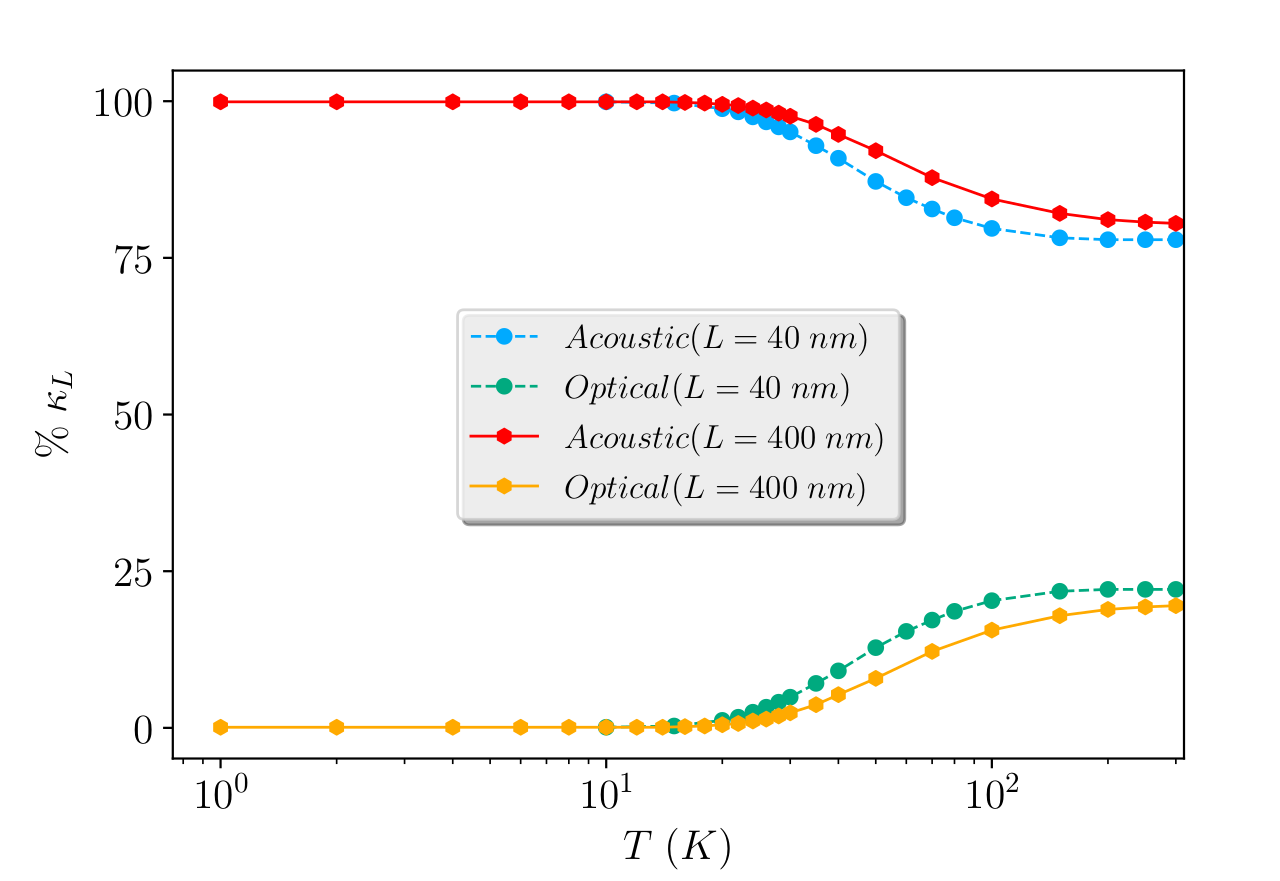}
    \caption{The variation of acoustic and optical mode contributions to total $\kappa_L$ in $\%$ with temperature for both $L$ = 40 nm and L = 400 nm.}
    \label{fig:AM_OM_contrib}
\end{figure}
where $\kappa_{L}$ ($\omega'$) is defined as \cite{Togo,Mizokami}
\begin{equation}
    \kappa_{L}(\omega') \equiv \frac{1}{NV_0} \sum_{\lambda} C_{\lambda} \textbf{v}_{\lambda} \otimes \textbf{v}_{\lambda}\tau_{\lambda} \delta (\omega'-\omega_{\lambda}) 
\end{equation}
with $\frac{1}{N}$ $\sum_{\lambda} \delta(\omega' - \omega_{\lambda})$ the weighted density of states (DOS). As realized from the detailed investigation of the phonon density of states and phonon dispersion relation in our earlier work \cite{kanka} on GeTe, a phonon frequency around $2.88$ THz can be found to be a good approximation of a separator between acoustic and optical modes \cite{kanka}. Figure \ref{fig:cumulativeAMOM} presents the cumulative lattice thermal conductivities along the $a$-axis ($\kappa_L^x$), the hexagonal $c$-axis ($\kappa_L^z$) and the average $\kappa_L^c$ as a function of phonon frequency for different temperatures for $L$ = 40 nm. The anisotropy of $\kappa_L$ for GeTe along the hexagonal $c$-axis and its perpendicular direction ($a$-axis) had been described in details elsewhere\cite{kanka}. The spectral representation of $\kappa_L^c$ indicates the density of heat carrying phonons with respect to the phonon frequencies and their contributions to $\kappa_L^c$. The density of modes goes to zero at a frequency where $\kappa_L^c$ reaches a plateau marking the separation between acoustic (frequency $<$ 2.87 THz) and optical (frequency $>$ 2.87 THz) modes. From Fig \ref{fig:cumulativeAMOM}, we note that below 50 K, the optical mode contribution gets very low and at 10 K it nearly vanishes. For understanding the role of boundary scattering, Fig \ref{fig:AM_OM_contrib} shows the relative contribution of acoustic and optical phonons to total $\kappa_L$ for $L$ = 40 nm and 400 nm. At higher temperature, we find that decreasing the boundary scattering effect by increasing the grain size from 40 nm to 400 nm can slightly enhance the contribution of acoustic modes from 77 $\%$ to around 80 $\%$, thereby reducing the optical mode contribution from 23 $\%$ to 20 $\%$. The contribution gradually increases (decreases) for acoustic modes (optical modes) and  below 20 K, the contribution saturates to almost 100 $\%$ for acoustic modes. We recall that, this vanishing contribution of optical modes can also be seen from Fig \ref{fig:LTC_mode_decomposed}.(c) and (d).
\\

\section{Phonon propagation length: Role of resistive processes for damping}{\label{section:Second sound}}

\begin{widetext}
\begin{figure}[H]
    \centering
    \includegraphics[width=1.0\textwidth]{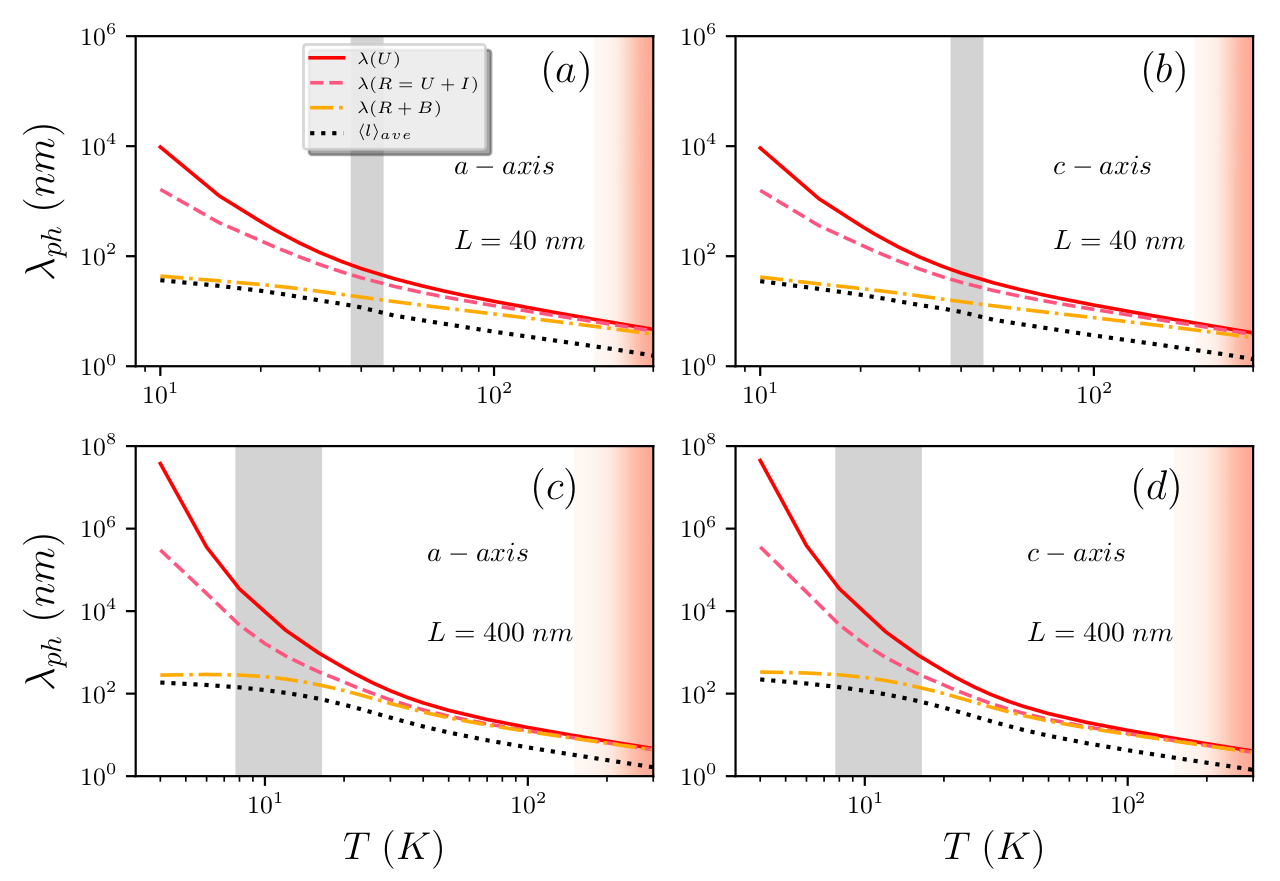}
    \caption{Phonon propagation length as a function of temperature for $L$ = 40 nm along (a) the $a$-axis and (b) the hexagonal $c$-axis of GeTe. Phonon propagation lengths for $L$ = 400 nm along the $a$-axis and the $c$-axis of GeTe are also shown in (c) and (d) respectively. Phonon propagation length due to various scattering processes are shown where $U$, $R$ (= $U$+$I$), $I$, $B$ denote umklapp, resistive, isotope and boundary scattering respectively. Variation of average mean free path ($\langle l \rangle_{ave}$) with temperature is also shown. Light red and gray shaded regimes indicate kinetic and hydrodynamic regime as defined in earlier sections.}
    \label{fig:second_sound_propagation}
\end{figure}
\end{widetext}

Second sound, a characteristic and important hydrodynamic heat transport phenomenon, refers to heat propagation as damped waves in a system \cite{Guyer1966_2, Cepellotti2015, umklapp_gang_2018}. This phenomenon is a direct manifestation of phonon collective motion due to the dominating contribution of normal scattering over the resistive scattering processes. For various materials, second sound in the Poiseuille flow regime had been identified at cryogenic temperature, both experimentally and theoretically \cite{Bi1972, Bi2018, Koreeda2007}. Following \cite{Bi2018}, we define two important quantities for investigation, namely second sound velocity or drift velocity ($\overline{v}$) and  phonon propagation length ($\lambda_{ph}$) as 
\begin{equation}
    \overline{v}_{j}^{2} = \frac{\sum_{\alpha} C_{\alpha}\mathbf{v}_{\alpha j}^{g}\cdot \mathbf{v}_{\alpha j}^{g}}{\sum_{\alpha}C_{\alpha}}
\end{equation}
and 
\begin{equation}
    \lambda_{ph} = \overline{v}/\langle \tau^{-1} \rangle _{ave}
\end{equation}
where, $C_{\alpha}$ is heat capacity of mode $\alpha$, $\mathbf{v}_{\alpha j}^{g}$ is phonon group velocity of mode $\alpha$ and $j$ can be either the component along the $a$-axis (x) or the hexagonal $c$-axis (z). We recall from our earlier study \cite{kanka} that the heat transfer of GeTe is anisotropic \cite{kanka, Campi2017}. Therefore, group velocities along the hexagonal $c$-axis and its perpendicular ($a$-axis) direction of GeTe \cite{kanka} are different, giving rise to different drift velocities and different phonon propagation lengths along these two directions. Figure \ref{fig:second_sound_propagation} shows the variation of second sound propagation length or the phonon propagation length with temperature along both $a$ and $c$-axis directions of GeTe. As $\lambda_{ph}$ is the distance that the phonon travels before damping \cite{Bi2018}, we present the separate contributions of different resistive processes for damping of a heat wave, namely $\lambda(U)$ (umklapp only), $\lambda(R)$ (resistive) and $\lambda(R+B)$ (resistive and boundary scattering) where  
\begin{equation}
    \lambda(U) = \overline{v}/\langle \tau_{U}^{-1} \rangle _{ave}
\end{equation}
\begin{equation}
    \lambda(R) = \overline{v}/\left(\langle \tau_{U}^{-1} \rangle _{ave}+\langle \tau_{I}^{-1} \rangle _{ave}\right)
\end{equation}
\begin{equation}
    \lambda(R+B) = \overline{v}/\left(\langle \tau_{U}^{-1} \rangle _{ave}+\langle \tau_{I}^{-1} \rangle _{ave}+\langle \tau_{B}^{-1} \rangle _{ave}\right)
\end{equation}
For comparison, average mean free path ($\langle l \rangle_{ave}$) has also been shown in Fig \ref{fig:second_sound_propagation}.

At higher temperature in the kinetic regime (light red shaded region in Fig \ref{fig:second_sound_propagation}) , $\lambda(U)$, $\lambda(R)$ and $\lambda(R+B)$ are found to be almost collapsed in a single curve for $L$ = 40 nm and $L$ = 400 nm grain sizes. This collapse is  

\onecolumngrid
due to the dominant contribution of umklapp scattering as the most significant resistive process for damping the phonon waves at higher temperatures. As temperature is lowered, the phonon propagation lengths due to different resistive processes start varying and are found to increase and separate out gradually. $\lambda(R+B)$ takes both resistive and boundary scattering into account for damping of phonon propagation and is found to approach to the phonon average mean free path ($\langle l \rangle_{ave}$) at very low temperature due to the significant boundary scattering in the ballistic regime. We observe that the damping by the umklapp and resistive scattering processes increases the phonon propagation length significantly compared to the $\langle l \rangle_{ave}$ of phonons, starting from the intermediate temperature. This arises due to the fact that $\langle l \rangle_{ave}$ is defined via uncorrelated phonon gas, and considers both normal and umklapp as resistive \cite{Bi2018}. Following the work of M. Markov \textit{et al.} \cite{Bi2018}, we note that at low temperature, the heat wave propagation length is close to the phonon propagation length calculated using only umklapp scattering as a damping source. For $L$ = 40nm (Fig \ref{fig:second_sound_propagation}.(a), (b)), in the region where Eq.\ref{eq:poiseuille} is satisfied (gray shaded region), propagation lengths, $\lambda(U)$ and $\lambda(R)$ are found to possess only slightly higher values ($\approx$ 6 times) compared to $\langle l \rangle_{ave}$. This indicates a feeble effect of phonon hydrodynamics for $L$ = 40 nm, consistent with the earlier description of phonon hydrodynamics from average scattering rates. On the other hand, $L$ = 400 nm displays a notable difference of the order of 10$^2$ and 10$^1$ between $\lambda(U)$ and $\langle l \rangle_{ave}$ and $\lambda(R)$ and $\langle l \rangle_{ave}$ respectively (Fig \ref{fig:second_sound_propagation}.(c) and (d)). Further, for $L$ = 400 nm, in the defined hydrodynamic regime (gray shaded region), $\lambda(U)$ reaches micron scale (from $\approx$ 1 $\mu$m at 16 K to $\approx$ 35 $\mu$m at 8 K), which further strengthens the possibility of observing second sound and phonon hydrodynamics for L = 400 nm.

We note here that experimentally, detecting second sound depends on the precise manifestation of many parameters. For example, the size of the experimental setup and the distance between pump and probe are two crucial parameters to realize the observation of second sound experimentally \cite{Cepellotti2015}.

\section{Thermal diffusivity: Role of various phonon scattering processes}{\label{section:Thermal diffusivity}}

\begin{figure}[H]
    \centering
    \includegraphics[width=0.5\textwidth]{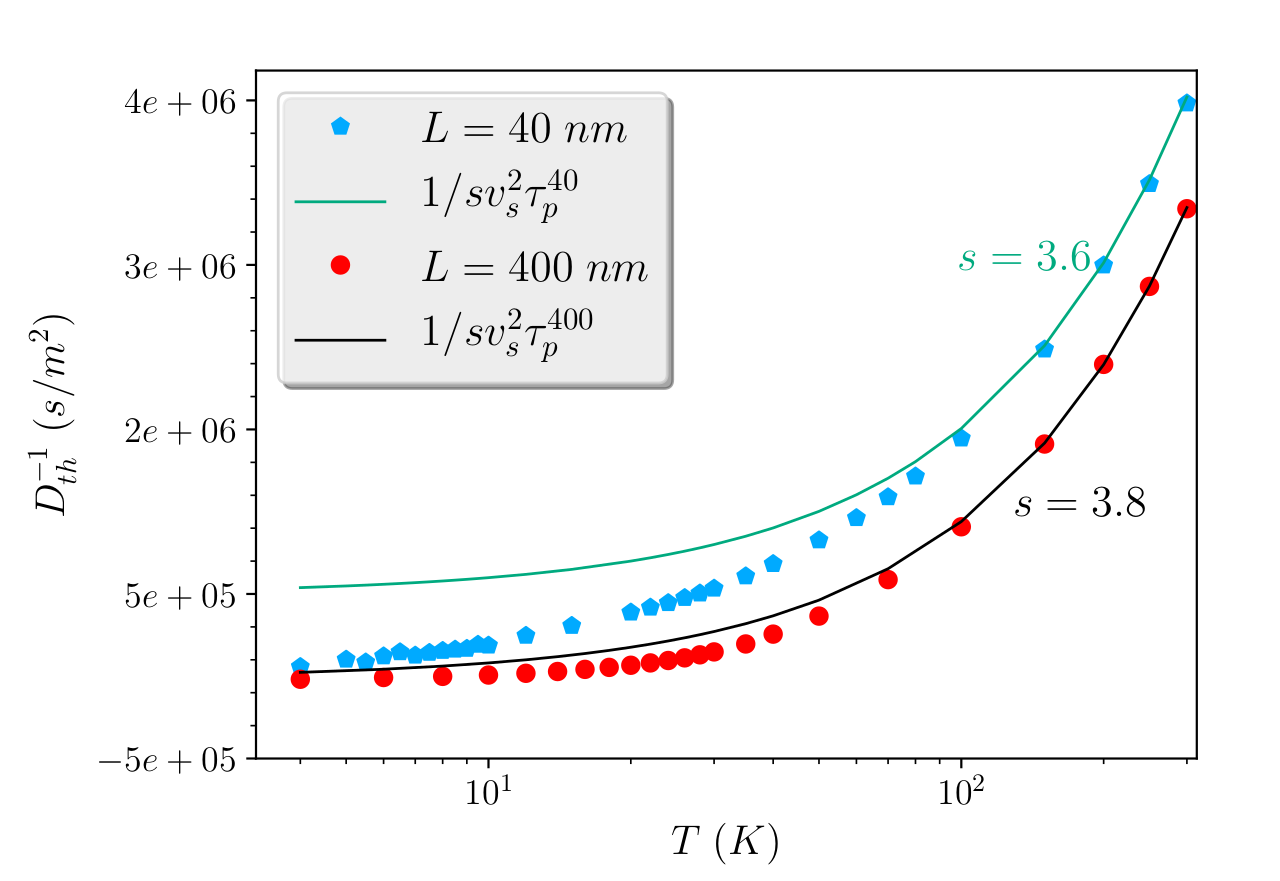}
    \caption{Variation of inverse thermal diffusivity ($D_{th}^{-1}$) with temperature for GeTe for $L$ = 40 nm and $L$ = 400 nm. For clear visualization temperature is shown in log-scale. Analytical expression of $D_{th}^{-1}$ = $1/sv_{s}^{2}\tau_{p}$ for both $L$ = 40 nm and $L$ = 400 nm are fitted at high temperature.}
    \label{fig:thermal diffusivity}
\end{figure}

Heat diffusion can be characterized by thermal diffusivity, an important quantity for inspection of the heat transfer mechanism in solids. It is defined as $D_{th}$ = $\kappa$/$\rho C$, where $\kappa$ is thermal conductivity, $\rho$ is mass density and $C$ is specific heat of the material, obtained through the heat equation via
\begin{equation}
    \frac{\partial T}{\partial t} - D_{th} \nabla^{2}T = 0
\end{equation}
Here $T$ and $t$ define temperature and time respectively. The thermal diffusion within a material describes the rate at which the heat flows or the speed of propagation of heat when a temperature gradient is introduced in the material \cite{Salazar_2003}. Therefore, higher thermal diffusivity quantifies the faster heat transfer. Generally, at high temperature, or specifically higher than the Debye temperature ($\Theta_D$ = 180 K for GeTe), $D_{th}$ decreases with $1/T$ due to the dominance of umklapp scattering between phonons \cite{Behnia_2019}. This can be simply understood from the fact that at higher temperature, $\kappa$ scales with $1/T$ and $C$ is almost constant, giving $D_{th}$ (= $\kappa$/$\rho C$) $\propto$ $1/T$. Figure \ref{fig:thermal diffusivity}
presents the inverse of thermal diffusivity as a function of temperature for crystalline GeTe. The trend shows a gradual decrement of $D_{th}^{-1}$ (increment of $D_{th}$) as the temperature is lowered.

Recently, considering heat carriers as diffusive quasi-particles, an universal boundary to thermal transport by phonons has been studied \cite{Behnia_2019, Mott-Ioffe-Regel}. At high temperature, it was found that $D_{th}$ exhibits a lower bound, governed by the sound speed in the material and a Planckian scattering time ($\tau_{p}$) via \cite{Behnia_2019, Mott-Ioffe-Regel, Strontium2018}
\begin{equation}
    D_{th} = sv_{s}^{2}\tau_{p}
\end{equation}
where, $\tau_p$ = $\frac{\hbar}{k_{B}T}$, $v_s$ is average sound speed and $s$ ($>$ 1) denotes a dimensionless parameter which is constant for a specific material. This bound had also been found for amorphous materials \cite{Behnia_2019}, which predicts a more fundamental quantum-mechanical origin to this phenomenon. 

We find a consistent behavior of $D_{th}^{-1}$ $\propto$ $T$ at higher temperature in the kinetic transport regime for both grain sizes (Fig \ref{fig:thermal diffusivity}), and fitting with the lower bound approximation using $v_s$ (= 1900 m/s \cite{Pereira}) and $\tau_p$ as known parameters for GeTe yields $s$ = 3.6 and 3.8 for $L$ = 40 nm and $L$ = 400 nm respectively.  
The closeness of these two values of $s$, or in other words, a nearly constant value of s for different grain sizes is representative of the fact that $s$ is constant for a particular material.

\begin{widetext}
\begin{figure}[H]
    \centering
    \includegraphics[width=1.0\textwidth]{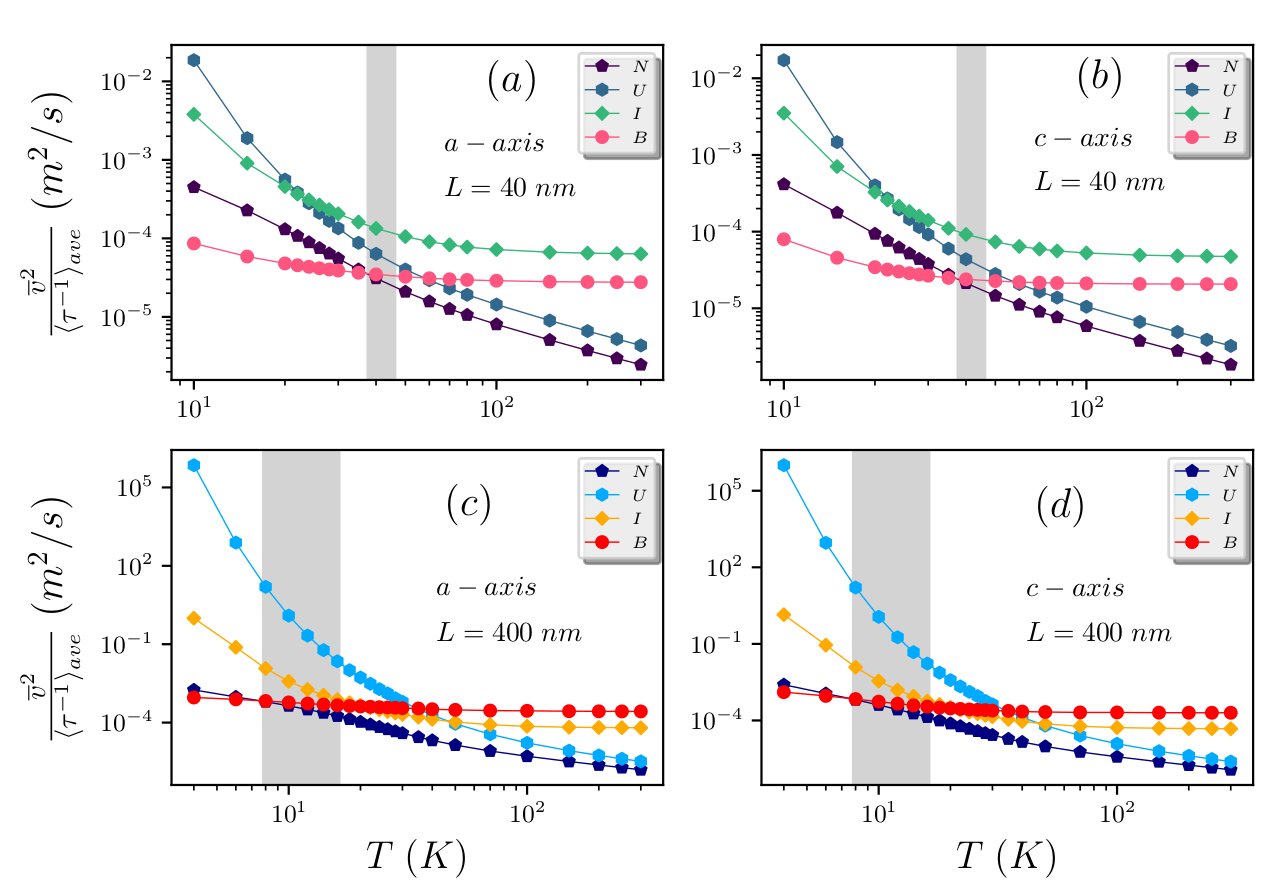}
    \caption{Thermal diffusivity in terms of $\overline{v}^2/\langle \tau^{-1} \rangle_{ave}$ is presented as a function of temperature for $L$ = 40 nm along (a) the $a$-axis and (b) the $c$-axis of crystalline GeTe. (c) and (d) present $\overline{v}^2/\langle \tau^{-1} \rangle_{ave}$ for $L$ = 400 nm along (c) the $a$-axis and (d) the $c$-axis of GeTe. $\langle \tau^{-1} \rangle_{ave}$ for normal, umklapp, isotope and boundary scattering are designated via $N$, $U$, $I$ and $B$ respectively.}
    \label{fig:thermal_diffusion_modes}
\end{figure}
\end{widetext}

To understand thermal diffusivity in terms of the contributions coming from different phonon scattering mechanisms, we perform a qualitative analysis for $D_{th}$. In general, $\kappa$ can be expressed as $\sum_{\lambda} C_{\lambda} v_{\lambda}^{2} / \tau_{\lambda}^{-1}$, which leads to $D_{th}$ $\approx$ $v^{2}/\tau^{-1}$. We take the thermodynamic averages of the numerator and denominator and calculate $\overline{v}^2/\langle \tau^{-1} \rangle_{ave}$ for different phonon scattering processes to qualitatively understand the essence of thermal diffusion in terms of these scattering mechanisms. Figure \ref{fig:thermal_diffusion_modes} shows $\overline{v}^2/\langle \tau^{-1} \rangle_{ave}$ as a function of temperature for $L$ = 40 nm and 400 nm along the $a$ and the $c$-axis of crystalline GeTe. For $L$ = 40 nm, at high temperature, phonon-isotope scattering is shown to contribute the maximum (Fig \ref{fig:thermal_diffusion_modes}.(a), (b)), whereas, for $L$ = 400 nm, phonon-boundary scattering shows the maximum contribution at high temperature (Fig \ref{fig:thermal_diffusion_modes}.(c), (d)). At very low temperature, umklapp scattering contributes the maximum for both grain sizes. While different scattering processes give comparable contributions to $D_{th}$ in the hydrodynamic regime (shaded region in Fig \ref{fig:thermal_diffusion_modes}.(a), (b)) for $L$ = 40 nm, umklapp scattering is shown to contribute several order higher values to $D_{th}$ for the hydrodynamic regime (shaded region in Fig \ref{fig:thermal_diffusion_modes}.(c), (d)) for $L$ = 400 nm compared to other scattering events. The very low resistive scattering processes, particularly the umklapp process (very high $\overline{v}^2/\langle \tau_{U}^{-1} \rangle_{ave}$) and correspondingly the very high normal scattering process (very low $\overline{v}^2/\langle \tau_{N}^{-1} \rangle_{ave}$) for $L$ = 400 nm, are found to be responsible for the enhancement of $D_{th}$ at low temperature. Therefore, the higher values of $\overline{v}^2/\langle \tau^{-1} \rangle_{ave}$ (or qualitatively $D_{th}$), for $L$ = 400 nm in the hydrodynamic regime, imply faster heat transfer,        

\onecolumngrid
which stems from the high normal scattering rate and low umklapp scattering rate in the denominator of $\overline{v}^2/\langle \tau^{-1} \rangle_{ave}$. This guarantees that strong momentum conserving phonon scattering will take place, which further supports the argument of hydrodynamic phonon flow for $L$ = 400 nm.

\section{The Kinetic-collective model predictions}{\label{section:KCM}}
In order to further investigate the detailed consequences of phonon hydrodynamics in GeTe, we employ the Kinetic-collective model (KCM) \cite{KCM-method2017} to detect and scrutinize the implications of phonon hydrodynamics from a different perspective. The KCM \cite{KCM-method2017} considers a part of the heat to be transferred via collective phonon modes, borne out of normal scattering events, apart from heat transfer by independent collisions. Therefore, lattice thermal conductivity can be expressed as a sum of both kinetic and collective contributions weighed by a switching factor ($\Sigma \in \left[0,1\right]$), which measures the relative weight of normal and resistive scattering processes \cite{KCM-method2017, Torres_2019}. While each mode possesses individual phonon relaxation time in the kinetic contribution term, the collective contribution is specified by an identical relaxation time for all modes \cite{alvarez2018thermalbook, KCM-method2017}. In the kinetic contribution term, the boundary scattering is included via the Matthiessen's rule as \begin{equation}
    \tau_{k}^{-1} = \tau_{U}^{-1} + \tau_{I}^{-1} + \tau_{B}^{-1}
\end{equation}
where $\tau_{k}$ is the total kinetic phonon relaxation time. On the contrary, a form factor $F$, calculated from the sample geometry, is used to incorporate boundary scattering in the collective term \cite{KCM-method2017, alvarez2018thermalbook}. The KCM equations are:
\begin{equation}
    \kappa_L = \kappa_{k} + \kappa_{c}
\end{equation}
\begin{equation}
    \kappa_k = (1-\Sigma)\int \hbar \omega \frac{\partial f}{\partial T}v^{2} \tau_{k} D \textit{d}\omega 
\end{equation}
\begin{equation}
    \kappa_c = (\Sigma F)\int \hbar \omega \frac{\partial f}{\partial T}v^{2} \tau_{c} D \textit{d}\omega 
\end{equation}
\onecolumngrid
\begin{widetext}
\begin{figure}[H]
    \centering
    \includegraphics[width=1.0\textwidth]{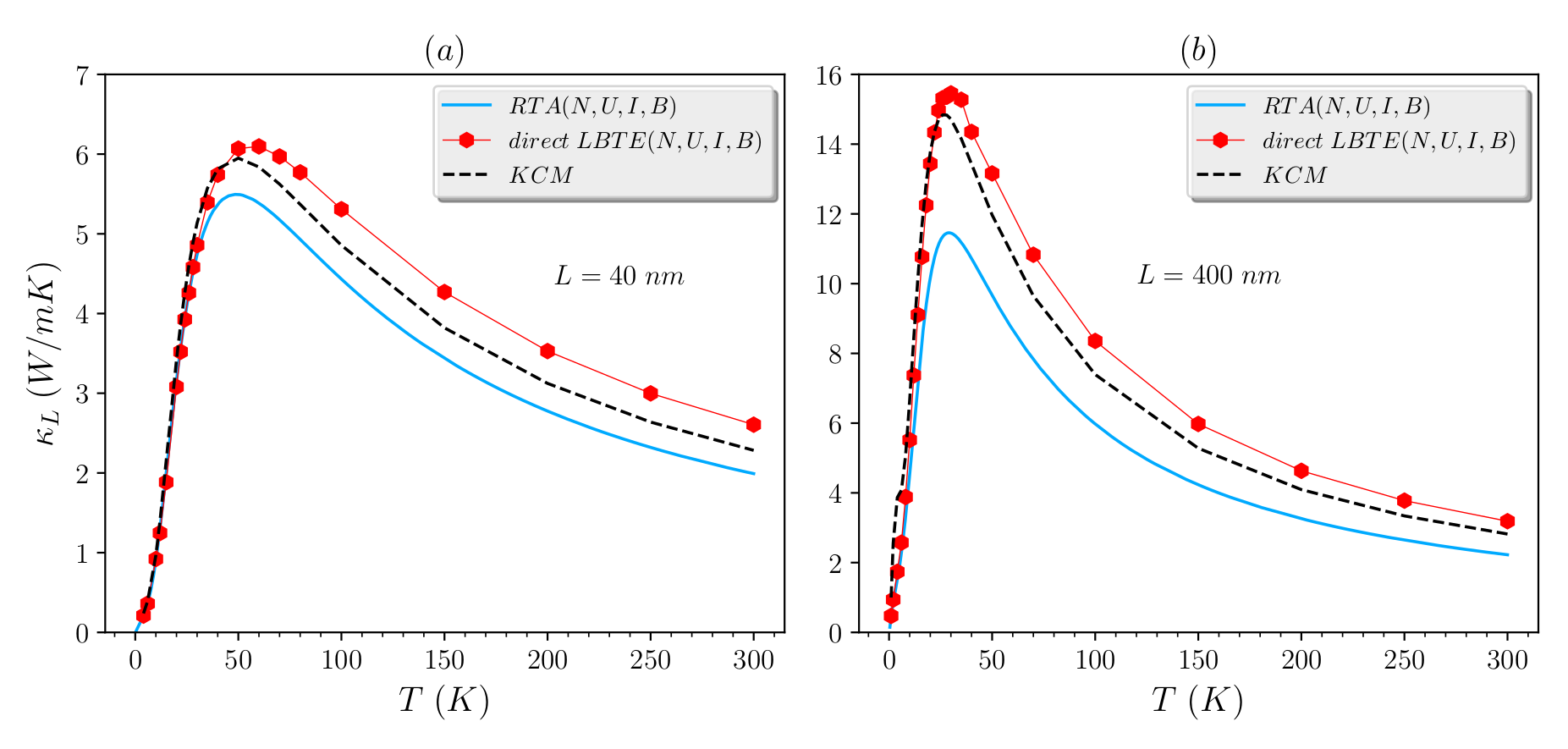}
    \caption{$\kappa_L$ of crystalline GeTe as a function of temperature for (a) $L$ = 40 nm and (b) $L$ = 400 nm grain sizes. The solutions for lattice thermal conductivity, obtained using direct solution of LBTE, RTA, and KCM are compared.}
    \label{fig:KCM_LBTE_RTA_compare}
\end{figure}
\vspace{-0.25cm}
\end{widetext}
\begin{equation}
    \Sigma = \frac{1}{1+\frac{\langle\tau_{N}\rangle}{\langle\tau_{RB}\rangle}}
\end{equation}

where $\kappa_k$ and $\kappa_c$ are kinetic and collective contributions to $\kappa_L$ respectively. $\langle\tau_{N}\rangle$ and $\langle\tau_{RB}\rangle$ designate average normal phonon lifetime and average resistive (considering $U$, $I$ and $B$) phonon lifetime respectively. $\langle \tau_{N} \rangle$ and $\langle \tau_{RB}\rangle$ are defined in the KCM \cite{KCM-method2017} as integrated mean free times:
\begin{equation}
    \langle \tau_{RB} \rangle = \frac{\int C_{1} \tau_k d\omega}{\int C_{1} d\omega}
\end{equation}
and 
\begin{equation}
    \langle \tau_{N} \rangle = \frac{\int C_{0} \tau_N d\omega}{\int C_{0} d\omega}
\end{equation}
where $\tau_k$ is the total kinetic relaxation time and phonon distribution function in the momentum space, represented in terms of $C_{i}(\omega)$, defined in \cite{KCM-method2017} as
\begin{equation}
    C_{i}(\omega) = \left( \frac{v |q|}{\omega}\right)^{2i} \hbar \omega \frac{\partial f}{\partial T} D
\end{equation}
where $v(\omega)$ is the phonon mode velocity and $\mid q\mid$ is modulus wave vector. $C_{0}$ represents the specific heat of mode $\omega$. f stands for Bose-Einstein distribution function, $v$ is mode velocity and $D(\omega)$ is phonon density of states for each mode. $\Sigma$ stands for the switching factor. $F$ is form factor approximated via \cite{alvarez2018thermalbook}
\begin{equation}
F(L_{eff}) = \frac{L_{eff}^2}{2\pi^2l^{2}} \left(\sqrt{1+\frac{4\pi^{2}l^2}{L_{eff}^2}}-1\right)   
\end{equation}
where, $L_{eff}$ is the effective length of the sample (in our system, we use $L_{eff}$ = $L$, the grain size) and $l$ is the characteristic non-local scale (details will be given later) \cite{Guyer1966_1, alvarez2018thermalbook}. $\tau_{c}$ denotes the total collective phonon relaxation time.
All the calculations regarding KCM have been done using the KCM.PY code \cite{KCM-method2017} with the PHONO3PY \cite{Togo} implementation.

As a first step, we seek to compare the results for $\kappa_L$ of GeTe, between direct solution of LBTE and that of the KCM \cite{KCM-method2017}. Figure \ref{fig:KCM_LBTE_RTA_compare} presents $\kappa_L$ as a function of temperature for both (a) $L$ = 40 nm and (b) $L$ = 400 nm, obtained using LBTE, RTA, and KCM. It is observed that at lower temperature, before the $\kappa_L$ peak, LBTE and KCM solutions are in excellent agreement. At higher temperature up to 300 K, a reasonably matching trend of $\kappa_L$ is retrieved using KCM, although exhibiting slightly lower values than the LBTE solutions. It can be noted that the LBTE solutions can be further lowered by incorporating the vacancy scattering in GeTe \cite{Campi2017, kanka} (we describe this vacancy effect later in this paper). It has been found that the experimental values of lattice thermal conductivity match quite well with KCM approximations for bulk Si, Ge, diamond and GaAs \cite{KCM-method2017, alvarez2018thermalbook}. As the study of this paper does not consist of experimental explorations at low temperature of GeTe, we are unable to comment whether KCM or direct-LBTE, matches well with experimental values for GeTe. However, we can further investigate the reason behind the differences between LBTE-direct solution and KCM predictions by closely studying the cumulative lattice thermal conductivity as a function of phonon frequency, obtained using both direct-LBTE and KCM at a temperature where 

\begin{figure}[H]
    \centering
    \includegraphics[width=0.5\textwidth]{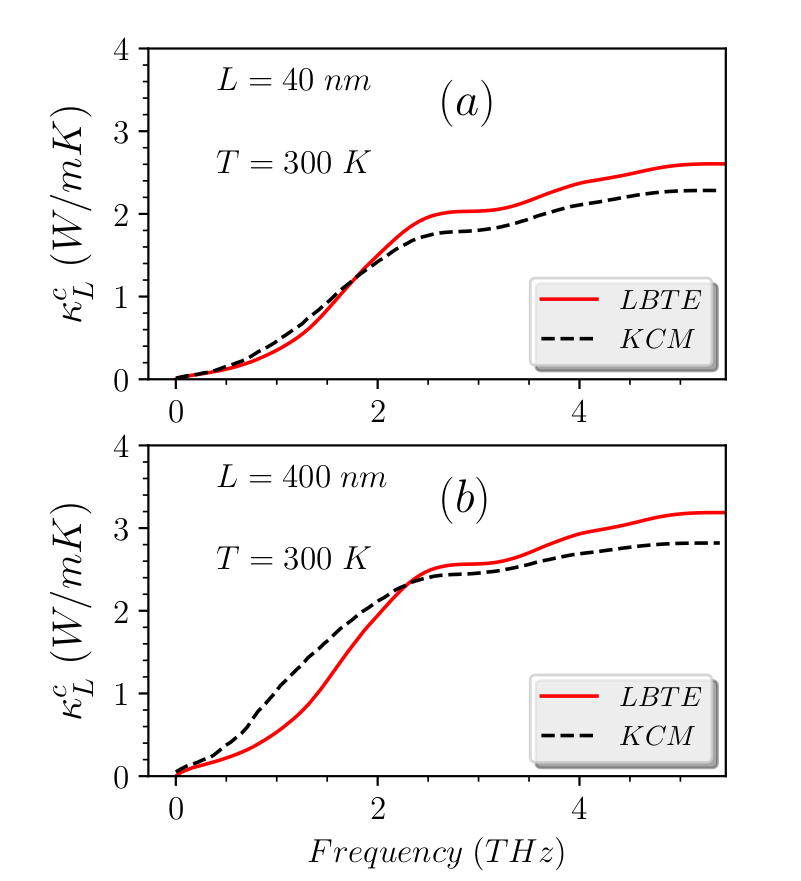}
    \caption{Cumulative lattice thermal conductivity ($\kappa_L^{c}$), obtained using both direct-LBTE and KCM, as a function of phonon frequency at 300 K for (a) $L$ = 40 nm and (b) $L$ = 400 nm.}
    \label{fig:LBTE_KCM_diff_cumulative}
\end{figure}
\noindent the difference is prominent. Figure \ref{fig:LBTE_KCM_diff_cumulative} shows a comparison between direct-LBTE and KCM cumulative lattice thermal conductivity ($\kappa_L^c$) of GeTe at 300 K for $L$ = 40 nm and $L$ = 400 nm. For $L$ = 40nm (Fig \ref{fig:LBTE_KCM_diff_cumulative}.(a)), we find that the difference between direct-LBTE and KCM solution is mostly appreciable in the optical modes regime (frequency $>$ 2.87 THz) of GeTe. For $L$ = 400 nm (Fig \ref{fig:LBTE_KCM_diff_cumulative}.(b)), although acoustic modes also show differences, a very feeble optical modes contribution is found to be responsible for restricting the $\kappa_L^{c}$ of KCM to a lower value than that of the direct solution of LBTE. A possible reason for the overestimation of direct-LBTE solutions was discussed by Feng $\textit{et al.}$ \cite{Tianli2017}, and it has been attributed to four-phonon scattering processes that can reduce the intrinsic thermal conductivity of solids, as shown for boron arsenide, Si and diamond. Also, Torres $\textit{et al.}$ \cite{Torres_2019} found similar discrepancies between LBTE and KCM solutions for MoS$_2$, borne out of the dissimilarities of these two solutions in the optical mode frequency regime. Therefore, we can predict a similar situation for GeTe and indicate that the four-phonon scattering, which can reduce the lifetime of the optical phonons, can be responsible for the slight overestimation of direct-LBTE solutions for GeTe compared to the KCM solutions beyond the peak of lattice thermal conductivity maximum in the temperature variation of $\kappa_L$. 

Also, as mentioned in \cite{alvarez2018thermalbook}, depending on the techniques for solving LBTE, the total relaxation time of the distribution function of the heat carrier can be different. Though direct-LBTE and KCM produce similar trends for relaxation times as both are the solutions borne out of the same phonon LBTE, there exists small differences between them, observed for Si and diamond \cite{alvarez2018thermalbook}, as KCM computes collective relaxation time through a switching factor $\Sigma$, which is different from the diagonalization of the full collision matrix as is done via the direct LBTE approach. However, in the proposed hydrodynamic temperature regime for GeTe as obtained earlier, the solutions of LBTE and KCM collapse satisfactorily. These observations mark KCM as a solid and reliable approach for our study on GeTe.

\subsection{The Kinetic and collective thermal transport}{\label{section:KCM1}}

Earlier in this paper, we established the strong and weak hydrodynamic effects of GeTe for grain sizes 400 nm and 40 nm respectively. To further scrutinize this effect, we calculate and show the contributions of the collective part ($\kappa_C$) of $\kappa_L$ to the total $\kappa_L$, using the KCM model for these two grain sizes. Starting from high temperature at 300 K, for both grain sizes, $\kappa_C$ is found to increase gradually as the temperature is lowered (Fig \ref{fig:KCM_collective_kinetic}.(a)). As expected, due to increased grain size, $L$ = 400 nm shows several order higher values of $\kappa_C$ compared to that of the $L$ = 40 nm case, as low temperature is being approached (Fig \ref{fig:KCM_collective_kinetic}.(a)). The hydrodynamic regime, calculated in earlier sections, is denoted via the shaded regions. For better realization, both collective ($\kappa_C$) and kinetic lattice thermal conductivities ($\kappa_{kin}$), obtained from KCM have been shown via Fig \ref{fig:KCM_collective_kinetic}. (a) and (b) respectively. Figure \ref{fig:KCM_collective_kinetic} (a) and (b) have been found to show the complementary behavior for $\kappa_C$ and $\kappa_{kin}$ respectively. The $\kappa_{kin}$ is observed to be the dominant part in determining the total lattice thermal conductivity, especially at low temperatures where $\kappa_{kin}$ shows a steep increase (Fig \ref{fig:KCM_collective_kinetic}.(b)). Moreover, in the hydrodynamic regime, the values of $\kappa_{kin}$ are $\approx$ 10 times and $\approx$ 10$^2$ times larger than the $\kappa_{C}$ for $L$ = 400 nm and 40 nm respectively. In a more illustrative way, Fig \ref{fig:KCM_collective_kinetic}.(c) presents the percentage contributions of $\kappa_C$ to the total KCM-lattice thermal conductivity. The complementary trends for $\kappa_{C}$ and $\kappa_{kin}$ are manifested in the smooth, monotonically decaying dependence of the percentage contribution of $\kappa_{C}$ to total lattice thermal conductivity on temperature (Fig \ref{fig:KCM_collective_kinetic}.(c)). Consistent with the earlier realizations of a weak hydrodynamic effect for smaller grain size, $L$ = 40 nm has been found to possess an extremely low percentage with negligible collective contribution in the designated hydrodynamic regime (gray shaded region) as shown in Fig \ref{fig:KCM_collective_kinetic}.(b). In contrast, we observe a substantial percentage of collective contribution to be present for $L$ = 400 nm, varying from $\approx$ 7$\%$ at 16 K to $\approx$ 18$\%$ at 8 K which validates the earlier approach of identifying a strong hydrodynamic regime for a larger grain size of GeTe.

\onecolumngrid
\begin{widetext}
\begin{figure}[H]
    \centering
    \includegraphics[width=1.0\textwidth]{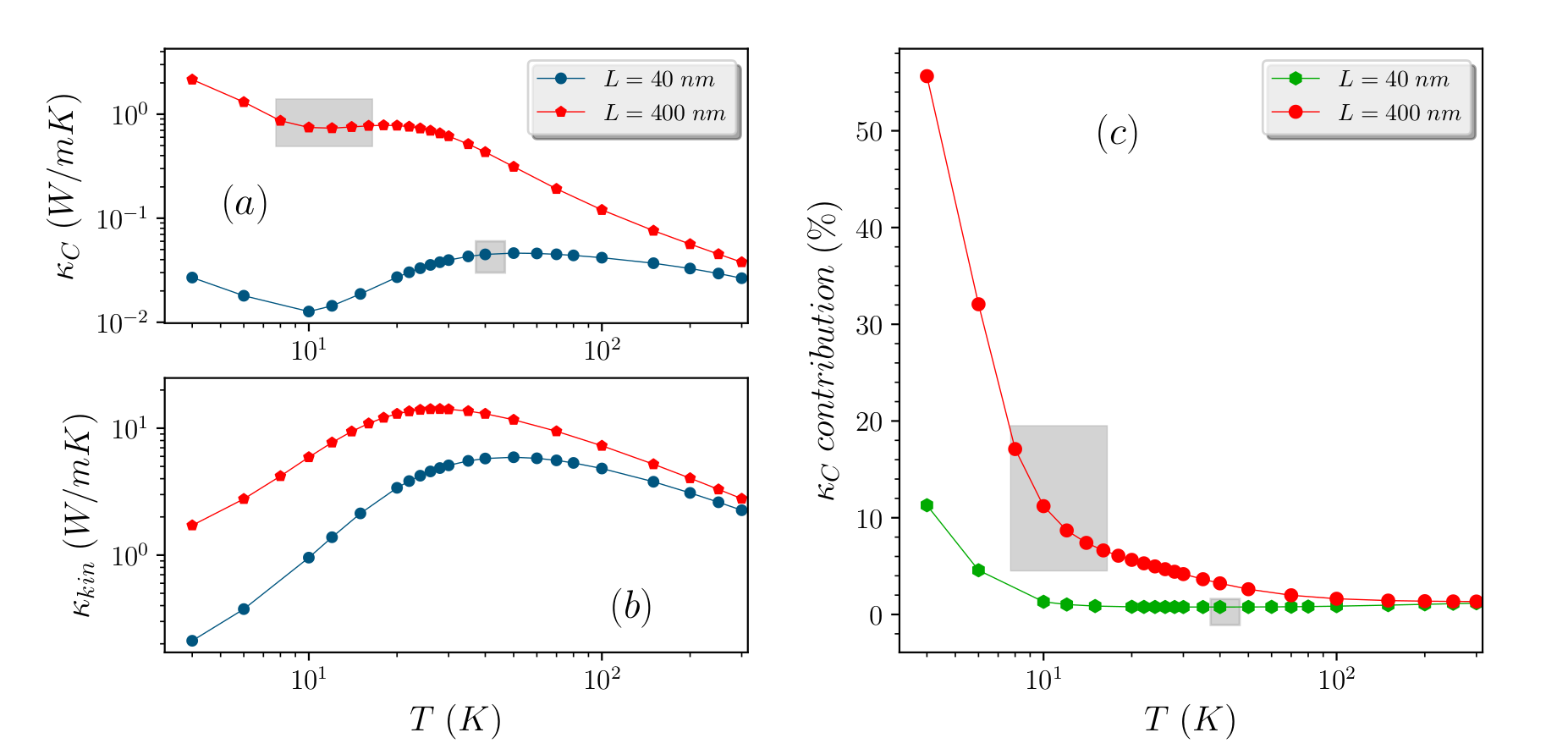}
    \caption{Variation of (a) collective ($\kappa_C$) and (b) kinetic lattice thermal conductivity ($\kappa_{kin}$) with temperature for $L$ = 40 nm and 400 nm. (c) The contribution of collective lattice thermal conductivity (in percentage) to the total lattice thermal conductivity is presented as a function of temperature for both $L$ = 40 nm and 400 nm.The hydrodynamic regimes under investigation are denoted via gray shaded zones in the $\kappa_C$.}
    \label{fig:KCM_collective_kinetic}
\end{figure}
\vspace{-0.3cm}
\end{widetext}

\begin{figure}[H]
    \centering
    \includegraphics[width=0.5\textwidth]{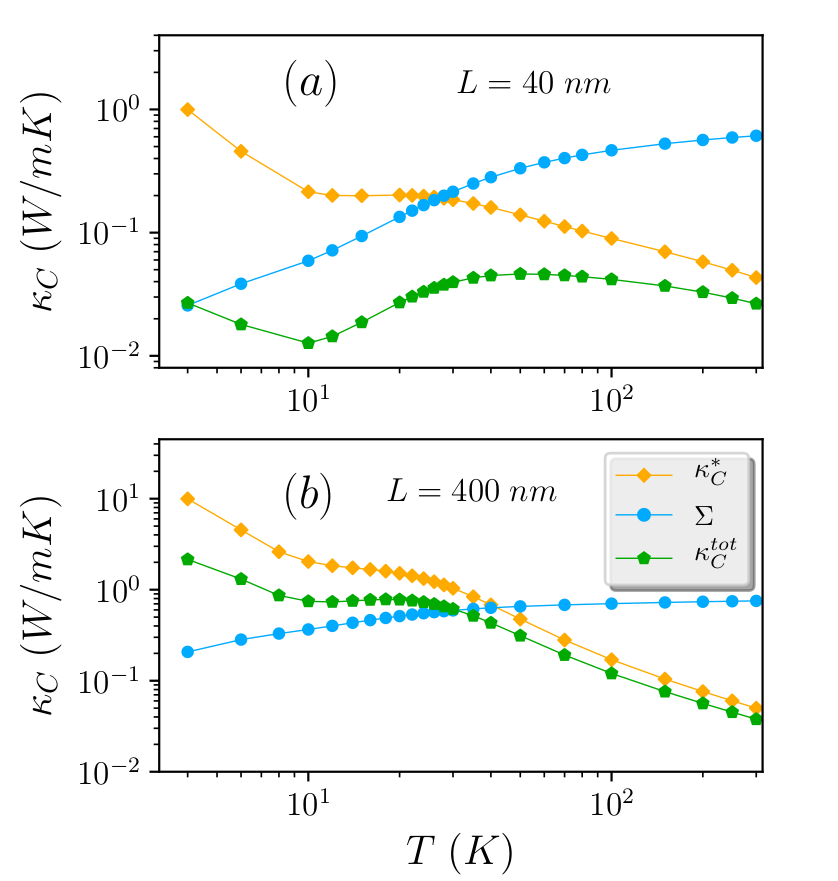}
    \caption{\label{fig:3:report}Variation of collective lattice thermal conductivity ($\kappa_C$) along with its constituent parameters $\Sigma$ and $\kappa_C^{*}$ with temperature for (a) $L$ = 40 nm and (b) $L$ = 400 nm.}
\end{figure}

Further, to investigate the minimum in the collective lattice thermal conductivity ($\kappa_C$), as shown in Fig \ref{fig:KCM_collective_kinetic}.(a), obtained from KCM, we present the variations of the constituent parameters of $\kappa_C$ with temperature. KCM represents the collective lattice thermal conductivity ($\kappa_C$) as \cite{KCM-method2017, alvarez2018thermalbook}
\begin{equation}
    \kappa_{C} = \Sigma F \int \hbar \omega \frac{\partial f}{\partial T} v^{2} \tau_{c} D d\omega = \kappa_{C}^{*} \Sigma
\end{equation}
Figure \ref{fig:3:report} shows the temperature variation of $\kappa_C$ along with the switching parameter ($\Sigma$) and $\kappa_C^{*}$ for $L$ = 40 nm (Fig \ref{fig:3:report}. (a)) and $L$ = 400 nm grain sized GeTe (Fig \ref{fig:3:report}. (b)). The minimum is found to originate as a result of the product between $\kappa_C^{*}$ and $\Sigma$. At temperatures close to 10 K, $\kappa_C^{*}$ exhibits a plateau-like regime and $\Sigma$ follows an increasing trend for both cases. Therefore, the multiplication of $\kappa_C^{*}$ and $\Sigma$ gives rise to the minima in $\kappa_{C}$. We note that $\Sigma_{L = 400}$ $>$ $\Sigma_{L = 40}$ for the whole temperature range and $\Sigma$ shows a steeper increasing trend for $L$ = 40 nm compared to $L$ = 400 nm, thus featuring a more prominent minimum in $\kappa_C$ for $L$ = 40 nm (Fig \ref{fig:3:report}. (a)). Therefore, the distinction of the collision matrix in terms of collective and kinetic relaxation times in the KCM approach is found to be the key for giving rise to the minimum in $\kappa_C$. 

\begin{widetext}
\begin{figure}[H]
    \centering
    \includegraphics[width=1.0\textwidth]{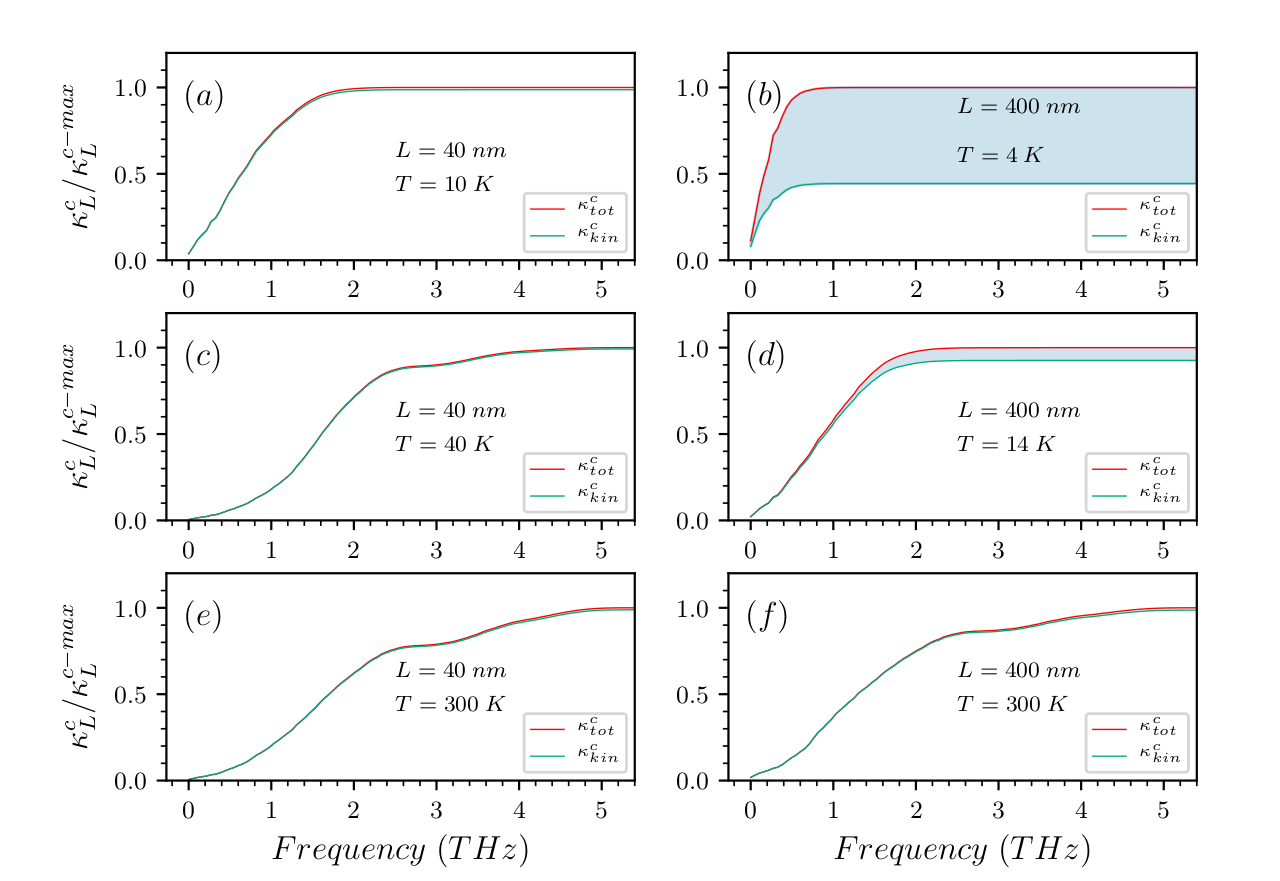}
    \caption{Cumulative lattice thermal conductivity ($\kappa_L^c$) using KCM method is presented as a function of phonon frequency. $\kappa_L^c$ for $L$ = 40 nm are shown for (a) $T$ = 10 K, (c) $T$ = 40 K and (e) $T$ = 300 K. $\kappa_L^c$ for $L$ = 400 nm are shown for (b) $T$ = 4 K, (d) $T$ = 14 K and (f) $T$ = 300 K. Total cumulative lattice thermal conductivity ($\kappa_{tot}^{c}$) and its kinetic contribution ($\kappa_{kin}^{c}$) are shown for each case. The difference between $\kappa_{tot}^{c}$ and $\kappa_{kin}^{c}$ is recognised as the collective contribution to $\kappa_L^c$ (prominently shown in (b) and (d) via shaded region). For each case, $\kappa_{tot}^{c}$ is divided by the maximum value of $\kappa_{tot}^{c}$ to modify the $y$-axis scale from 0 to 1.}
    \label{fig:KCM_cumulative}
\end{figure}
\end{widetext}

An alternative way to understand the collective contribution is to present the cumulative lattice thermal conductivity ($\kappa_L^c$), obtained from KCM, as a function of phonon frequency. Figure \ref{fig:KCM_cumulative} displays the variation of total ($\kappa_{tot}^{c}$) and kinetic lattice thermal conductivity ($\kappa_{kin}^{c}$) in a cumulative way with temperature. For, $L$ = 40 nm, as temperature is lowered from 300 K to 10 K  

\onecolumngrid

\noindent (Fig \ref{fig:KCM_cumulative}.(e), (c) ,(a)), no substantial difference is observed between $\kappa_{tot}^{c}$ and $\kappa_{kin}^{c}$, implying the negligible contribution of the collective part of $\kappa_L^c$. As temperature is lowered from 300 K to 4 K (Fig \ref{fig:KCM_cumulative}.(f), (d), (b)), for $L$ = 400 nm, a substantial difference is found to develop between $\kappa_{tot}^{c}$ and $\kappa_{kin}^{c}$, indicating a gradual increment of the contribution coming from the collective part of $\kappa_L^c$ (shown via the shaded region in Fig \ref{fig:KCM_cumulative}.(b), (d)). Figure \ref{fig:KCM_cumulative}.(b) and (d) also feature a crucial frequency dependence. At T = 14 K (Fig \ref{fig:KCM_cumulative}.(d)), we observe that the difference between $\kappa_{tot}^{c}$ and $\kappa_{kin}^{c}$ pops out in the acoustic regime (defined as frequency $<$ 2.87 THz \cite{kanka}) and then stays constant thereafter. No contribution is found to come from optical modes (defined as frequency $>$ 2.87 THz) in the hydrodynamic regime of GeTe. A similar feature can be found at $T$ = 4 K for $L$ = 400 nm (Fig \ref{fig:KCM_cumulative}.(b)).

\subsection{Hydrodynamic KCM and Knudsen number}{\label{section:KCM2}}

Instead of dealing separately with kinetic effects in the kinetic transport regime and hydrodynamic derived conditions in the collective regime, it is often helpful to envisage the thermal transport through a full hydrodynamic description, where both kinetic and hydrodynamic limits can be achieved under certain conditions \cite{alvarez2018thermalbook}. This generalized equation, which is an extension of the Guyer and Krumhansl equation \cite{Guyer1966_1} done in the KCM framework \cite{alvarez2018thermalbook}, named the hydrodynamic KCM equation, reads:
\begin{equation}
    \tau \frac{\textit{d}\textbf{Q}}{\textit{dt}}+ \textbf{Q} = -\kappa \nabla T + l^{2}\left( \nabla^{2}\textbf{Q} + 2\nabla \nabla \cdot \textbf{Q} \right)
\end{equation}
where $\tau$ is the total phonon relaxation time, $\textbf{Q}$ is the heat flux, $\kappa$ is phonon thermal conductivity, and $l$ is the non-local length. We investigate this non-local length ($l$) that determines the non-local range in phonon transport. The generalized form of non-local length is \cite{alvarez2018thermalbook}:
\begin{equation}
    l^{2} = \hat{l}_{K}^{2} \cdot \left(1-\Sigma\right) + \hat{l}_{C}^{2} \cdot \Sigma = l_{K}^{2} + l_{C}^{2}
\end{equation}
Here, the hat $\hat{}$ defines the limit situation (either kinetic limit or collective limit), and $l_{K}$ and $l_{C}$ define non-local length for kinetic and collective limit respectively.

For a clear demonstration of the hydrodynamic regime of GeTe, within the hydrodynamic KCM framework, it is instructive to compute the Knudsen number, defined via total non-local length ($l$) as
\begin{equation}
    Kn = l/L
\end{equation}
where, $L$ is the grain size for GeTe. Ideally, the Fourier law is seen to be recovered for low Kn values, whereas hydrodynamic behavior becomes important when Kn gets higher \cite{alvarez2018thermalbook, GUO20151}.


Till now, all the findings in this paper have indicated a significant hydrodynamic effect with a prominent low temperature range in GeTe for the larger grain size ($L$ = 400 nm), compared to the smaller grain size ($L$ = 40 nm), where the effect is seen to be weak. Therefore, to further confirm our findings in a quantitative way, the Knudsen number Kn is calculated and presented as a function of temperature for GeTe with grain size $L$ = 400 nm in Fig \ref{fig:KCM_knudsen}. Starting from 300 K, a gradual increasing trend of Kn is observed as the temperature is lowered. Holding the conceptual similarities with fluid flow or more specifically micro-scale gas flow, the Knudsen number (Kn) has been described in earlier heat transfer studies \cite{GUO20151, Bi2018} to identify a phonon hydrodynamic regime when 0.1 $\leq$ Kn $\leq$ 10. Remarkably, we find that the hydrodynamic regime for GeTe for $L$ = 400 nm (shown via the gray shaded region in Fig \ref{fig:KCM_knudsen}), obtained and identified using various methods in earlier sections, falls under the regime of 0.1 $\leq$ Kn $\leq$ 10, which is in agreement with the criteria for phonon hydrodynamics. From Fig \ref{fig:KCM_knudsen}, it is also observed that all the points that lie inside the range 0.1 $\leq$ Kn $\leq$ 10 (18 K $<$ T $<$ 50 K) do not necessarily fall within the hydrodynamic regime defined using Eq. \ref{eq:hydro} and Eq. \ref{eq:poiseuille} (gray shaded region). We recall from Fig \ref{fig:average_phonon_scattering}.(b) that this region corresponds to the condition $\langle \tau_{N}^{-1}\rangle_{ave}$ $>$ $\langle \tau_{R}^{-1}\rangle_{ave}$ $>$ $\langle \tau_{B}^{-1}\rangle_{ave}$, which can be referred to as the Ziman hydrodynamic regime \cite{Cepellotti2015}. Therefore, using Kn and an average scattering rate comparison, both the Poiseuille hydrodynamic regime and the Ziman hydrodynamic regime can be realized for GeTe. However, we mention here that although this regime follows the prescribed hierarchy of the Ziman hydrodynamic conditions, the differences between the scattering rates are found to be quite small.

From Fig \ref{fig:KCM_knudsen}, we also clearly distinguish the kinetic transport regime from the temperature variation of the Knudsen number as the region that satisfies Kn $\leq$ 0.1. Moreover, the region corresponding to Kn $\geq$ 10 (Fig \ref{fig:KCM_knudsen}), commonly understood as a free molecular flow or ballistic regime in fluid hydrodynamics, consistently corresponds to the ballistic thermal transport regime \cite{GUO20151}. Thus, a hydrodynamic KCM study of GeTe, realized through characteristic non-local length and Knudsen number estimation, provides a quantitative picture that agrees well with the scattering rate analysis and is consistent with various thermal transport property calculations in this study. 
\onecolumngrid
\begin{widetext}
\vspace{-0.6cm}
\begin{figure}[H]
    \centering
    \includegraphics[width=0.65\textwidth]{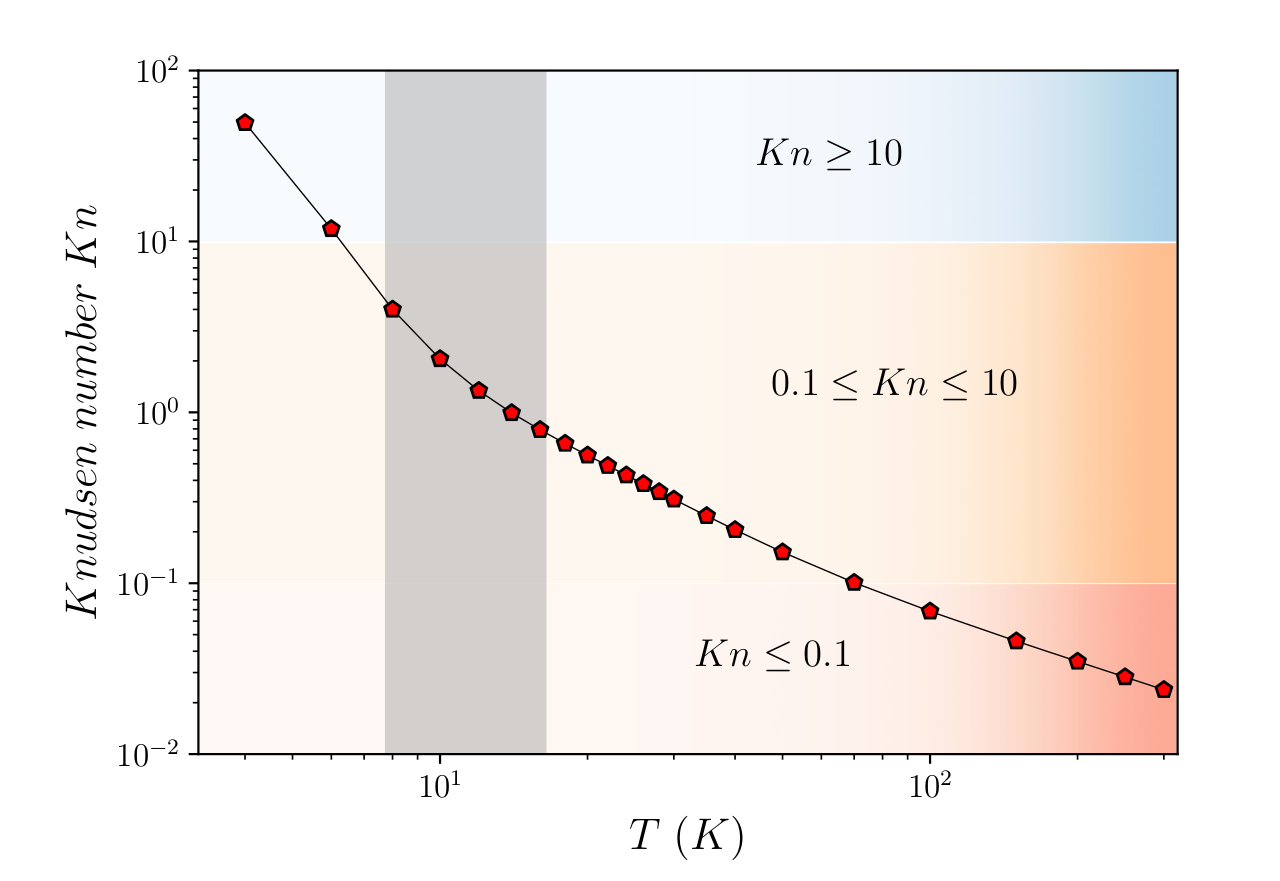}
    \caption{The variation of Knudsen number (Kn = $l/L$) with temperature for GeTe with $L$ = 400 nm. Light red, light orange and light blue define kinetic, hydrodynamic and ballistic regimes respectively (see text). The gray shaded region indicates hydrodynamic regime calculated from Eq.\ref{eq:hydro} and Eq.\ref{eq:poiseuille}.}
    \label{fig:KCM_knudsen}
\end{figure}
\end{widetext}

\onecolumngrid
\begin{widetext}
\begin{figure}[H]
    \centering
    \includegraphics[width=1.0\textwidth]{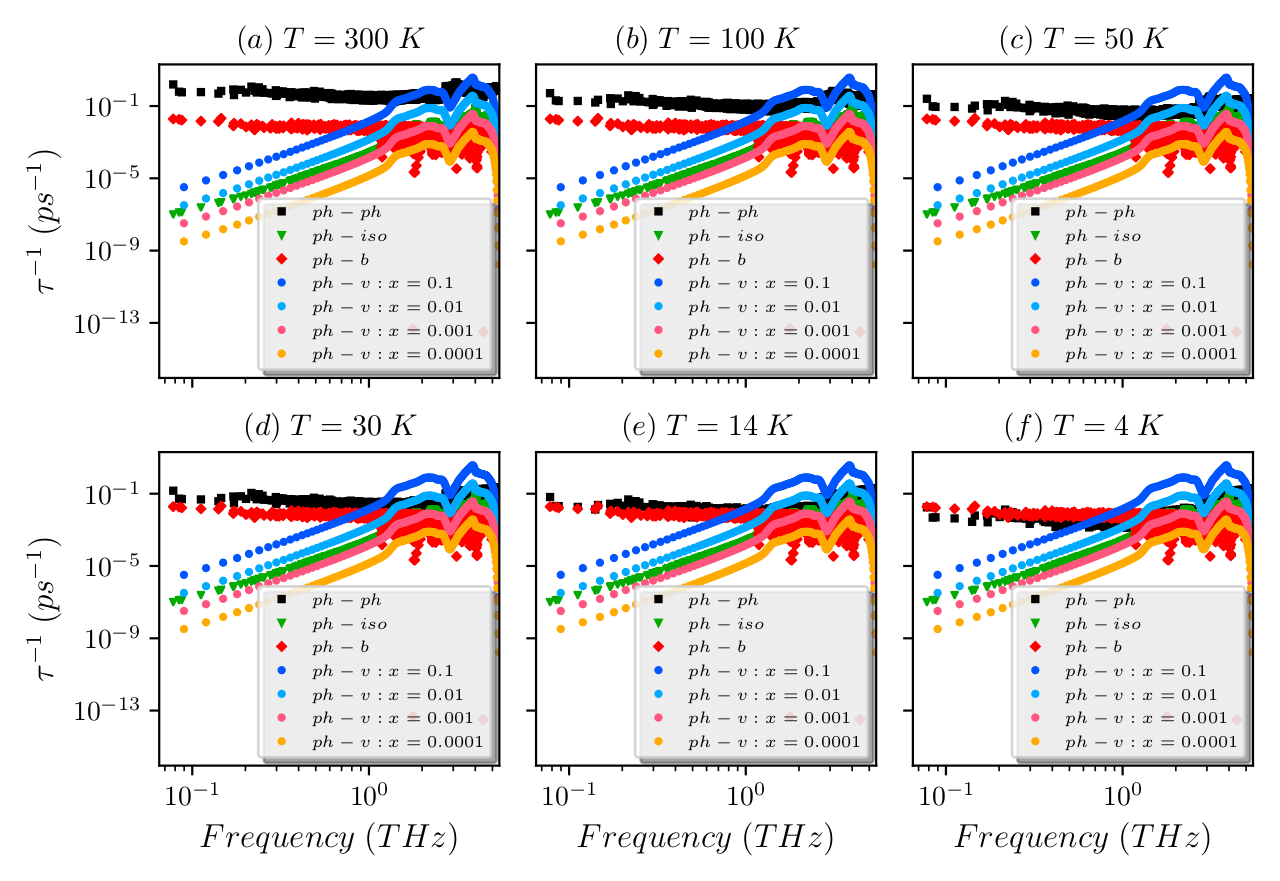}
    \caption{Phonon scattering rates are shown as a function of phonon frequency for crystalline GeTe for various temperatures: (a) $T$ = 300 K, (b) $T$ = 100 K, (c) $T$ = 50 K, (d) $T$ = 30 K, (e) $T$ = 14 K and (f) $T$ = 4 K. Phonon-phonon, phonon-isotope, phonon-boundary and phonon-vacancy scattering rates (for four different vacancy concentrations in percentage: $x$ = 0.0001, 0.001, 0.01 and 0.1) are shown.}
    \label{fig:ph-scattering_L_0_4}
\end{figure}
\end{widetext}

\section{Phonon-vacancy scattering: Effect of vacancy on hydrodynamic regime}{\label{section:vacancy}}

In the previous sections, we calculated the phonon lifetime and consequently the lattice thermal conductivity using phonon-phonon, phonon-isotope, and phonon-boundary scattering processes. However, the role of vacancies in the thermal transport is an important consideration. For GeTe, phonon-vacancy scattering has been found to be crucial to accurately describe the experimental data through the theoretical calculations at room temperature \cite{Campi2017} as well as at high temperatures \cite{kanka}. In our earlier work \cite{kanka}, the hole concentration of GeTe was found to be  6.24 $\times$ 10$^{19}$ cm$^{-3}$, indicating a vacancy concentration ($x$) of $\approx$ 0.08 $\%$. To understand the effect of phonon-vacancy scattering for GeTe, compared to the other phonon scattering events, scattering rates have been calculated and shown in Fig \ref{fig:ph-scattering_L_0_4} for grain size ($L$) of 400 nm as a function of frequency for different temperatures. The phonon scattering rates by vacancy defects are calculated following the work by Ratsifaritana $\textit{et al.}$ \cite{ratsifaritana} as
\begin{equation} {\label{eqvacancy}}
    \frac{1}{\tau_{V}(\omega)} = x \left(\frac{\Delta M}{M}\right)^{2}\frac{\pi}{2}\frac{\omega^{2}g(\omega)}{G'}
\end{equation}
where, $x$ is the density of vacancies or vacancy concentration, $G'$ denotes the number of atoms in the crystal, and $g(\omega)$ is the phonon density of states (PDOS). Using vacancies as isotope impurity, 
Ratsifaritana $\textit{et al.}$ \cite{ratsifaritana} evaluated mass change $\Delta M$ = 3 M, where M is the mass of the removed atom. It is noted from Eq.\ref{eqvacancy} that the phonon-vacancy relaxation time is temperature independent. Figure \ref{fig:ph-scattering_L_0_4} shows that the effect of phonon-vacancy scattering is quite significant at the high frequency regime. As we gradually lower the temperature from 300 K to 4 K, phonon-phonon scattering gradually decreases and at very low temperature phonon-boundary scattering overpowers the phonon-phonon scattering  
\begin{figure}[H]
    \centering
    \includegraphics[width=0.5\textwidth]{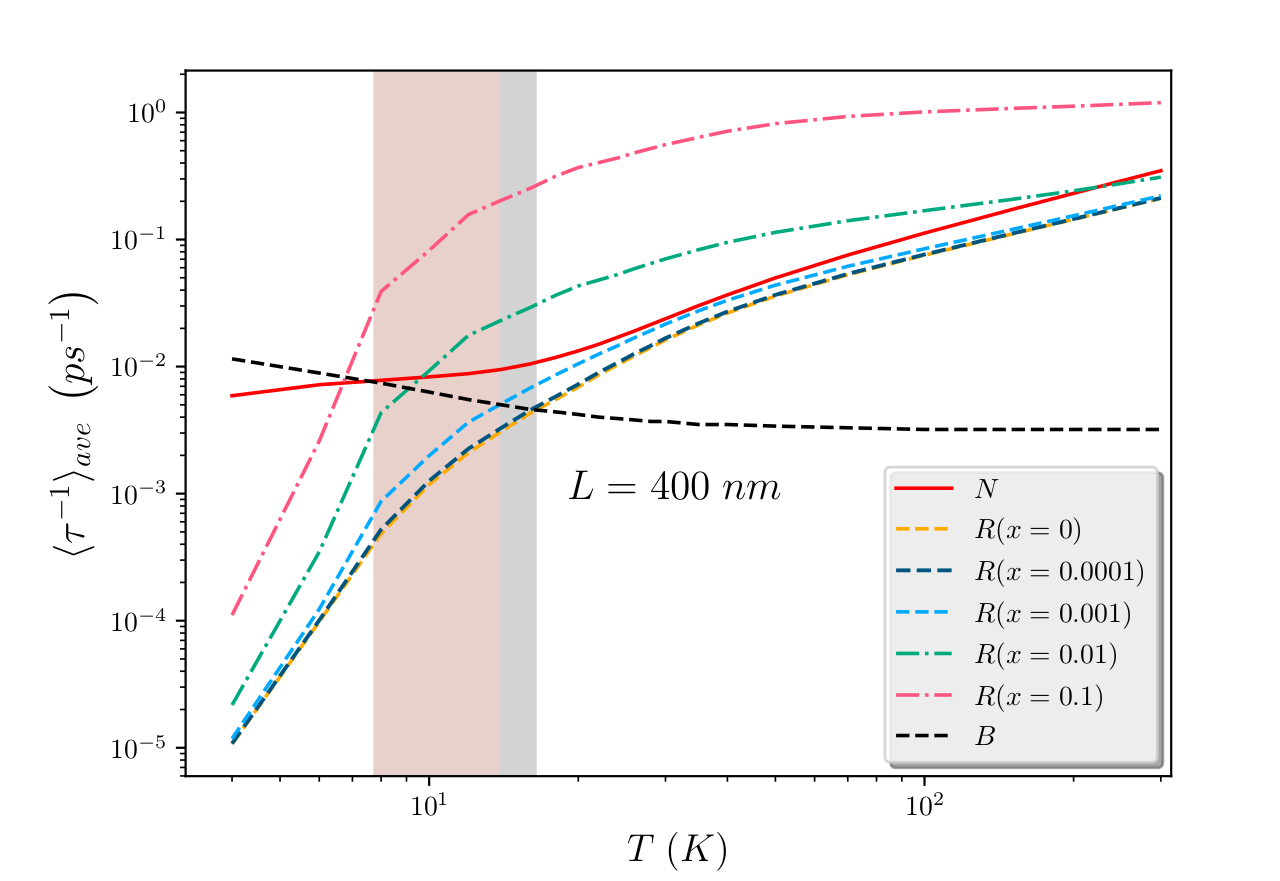}
    \caption{Thermodynamic average phonon scattering rates as a function of temperature in log-log scale for GeTe for grain size ($L$) = 400 nm. $N$, $R$ and $B$ denote normal, resistive and boundary scattering respectively. Here $R$ = $U$ + $I$ + $V$, where $U$, $I$ and $V$ stand for umklapp, phonon-isotope and phonon-vacancy scattering respectively. Four vacancy concentrations are used: $x$ = 0.0001, 0.001, 0.01, and 0.1. The gray shaded region corresponds to the hydrodynamic regime for $x$ = 0 and 0.0001. The light pink shaded region defines the hydrodynamic regime for $x$ = 0.001.}
    \label{fig:ave_scat_vacancy}
\end{figure}
(Fig \ref{fig:ph-scattering_L_0_4}.(f)). Therefore, the hydrodynamic regime can be significantly modified depending on the order of vacancy concentration, which adds up to the phonon resistive scattering. The average phonon scattering rates are investigated (Fig \ref{fig:ave_scat_vacancy}) incorporating phonon-vacancy scattering as resistive scattering along with umklapp and phonon-isotope scattering, to observe the effect on the previously defined hydrodynamic regime for $L$ = 400 nm. Starting from a pure GeTe ($x$ = 0), four different vacancy concentrations are considered: $x$ = 0.0001, $x$ = 0.001, $x$ =0.01, and $x$ = 0.1. Figure \ref{fig:ave_scat_vacancy} shows that $x$ = 0.0001 gives negligible impact on the modification of the hydrodynamic window, whereas $x$ = 0.001 slightly alters the regime by shrinking the window from $\approx$ 8-16 K (gray shaded region) to $\approx$ 8-14 K (light pink shaded region). Further increasing $x$ ($>$ 0.001) has been shown to affect the hydrodynamic window for GeTe drastically. Strong vacancy scattering effects coming from $x$ = 0.01 and 0.1 seem to completely shrink and vanish the hydrodynamic regime present in GeTe. Looking at Fig \ref{fig:ph-scattering_L_0_4}, we note that the phonon-isotope scattering rate (green symbols) separates these two kinds of vacancy scattering. Scattering rates corresponding to $x$ = 0.0001 and 0.001 lie below the phonon-isotope scattering rate while $x$ = 0.01 and 0.1 lie above it. Thus, phonon-isotope scattering ($\tau_{I}^{-1}(\omega)$) acts as an indicator for phonon vacancy scattering rates ($\tau_{V}^{-1}$) for controlling the hydrodynamic regime for GeTe. When $\tau_{I}^{-1}(\omega)$ $>$ $\tau_{V}^{-1}(\omega)$, the hydrodynamic regime is found to exist whereas the condition $\tau_{I}^{-1}(\omega)$ $<$ $\tau_{V}^{-1}(\omega)$ is responsible for shrinking and vanishing of the hydrodynamic window.

\section{Summary and conclusions}{\label{section:summary}}

A systematic and in-depth theoretical investigation has been carried out to understand the low temperature thermal transport in low $\kappa_L$ chalcogenide material GeTe in a crystalline phase. The low-temperature investigation reveals a plethora of novel and interesting phenomena related to phonon scattering that helps us to attain a complete understanding of the different competitive phonon scattering mechanisms and their implications. Lattice dynamics simulations have been carried out using density functional methods and solving linearized Boltzmann transport equations for a wide temperature range, starting from room temperature (300 K) to as low as 4 K, for GeTe. Two different grain sizes are considered to investigate the role of phonon-boundary scattering. Normal, umklapp, phonon-isotope and phonon-boundary scattering are separately distinguished and the thermodynamic average scattering rates are studied as a function of temperature. A prominent hydrodynamic regime is found for $L$ = 400 nm grain size, which gets weak while increasing the phonon-boundary scattering introducing a smaller grain size ($L$ = 40 nm). The variations of lattice thermal conductivity ($\kappa_L$) are studied, and comparing direct LBTE solutions with the single-mode relaxation time (RTA) approximations further shows the signatures of kinetic, hydrodynamic, and ballistic heat transport regimes of GeTe. Mode-wise decomposition of $\kappa_L$ shows the dominant heat transfer by acoustic phonons, which even increases its contribution upon increasing the grain size. Different acoustic modes (TA1, TA2, LA) are shown to evolve in a different way as a function of temperature. The transverse acoustic mode TA1 shows the maximum contribution throughout the temperature range, while at low temperature the contribution even reaches 80 $\%$ for the total acoustic $\kappa_L$. On the other hand, TA2 and LA modes contribute $\approx$ 20 $\%$ and 0 $\%$, respectively, at extreme low temperatures. Second sound propagation lengths have been calculated using various resistive processes as the damping sources and compared with the average mean free path of phonons. For larger grain size, the phonon propagation length corresponds to the umklapp and resistive damping, reaching up to the micron scale. Heat diffusion of GeTe has also been characterized using thermal diffusivity. At high temperature, the universal lower bound of thermal diffusivity has been found to exist, and it is governed by the sound speed in the material and the Planckian scattering time. The parameter s has been identified for GeTe and found to be around $3.6-3.8$. We perform a qualitative analysis for $D_{th}$ by calculating $\overline{v}^2/\langle \tau^{-1} \rangle_{ave}$ for different phonon scattering processes to understand the thermal diffusion in terms of various scattering mechanisms. Whereas almost comparable contributions are found for the $D_{th}$ for $L$ = 40 nm, extremely low umklapp scattering is shown to contribute several order higher values to $D_{th}$ compared to other scattering processes in the hydrodynamic regime for $L$ = 400 nm. The higher values of $\overline{v}^2/\langle \tau^{-1} \rangle_{ave}$, for $L$ = 400 nm in the hydrodynamic regime, indicate faster heat transfer which comes from the simultaneous high normal scattering and low umklapp scattering rate. Thus a strong momentum conserving phonon scattering occurs, which further supports the possibility of hydrodynamic phonon flow for $L$ = 400 nm. 

The kinetic-collective model (KCM) has also been implemented to scrutinize the hydrodynamic behavior of GeTe  from a different perspective. Collective and kinetic contributions to the thermal transport properties are understood via a switching  factor, which measures the relative weight of normal and resistive scattering. The KCM predictions on lattice thermal conductivity ($\kappa_L$) match quite well with the direct solution of LBTE, especially at low temperatures below the $\kappa_L$ peak. This leads us to calculate the collective contribution to the KCM-$\kappa_L$, and up to a contribution of 18$\%$ is found to exist for larger grain size (400 nm) in the hydrodynamic regime of GeTe. The characteristic non-local length, an indicator of the non-local range for phonon transport, along with the grain size, gives the Knudsen number (Kn), which further quantifies and validates the various thermal transport regimes, namely kinetic, hydrodynamic, and ballistic regimes. Finally, phonon-vacancy scattering for GeTe is incorporated considering various vacancy concentrations ranging from $x$ = 0.0001 $\%$ to 0.1 $\%$. For $x$ $>$ 0.001, vacancies are found to contribute significantly to the total resistive scattering and alter the hydrodynamic window severely. Thus, a proper combination of vacancy concentration and grain size emerged as important controlling parameters to observe phonon hydrodynamics in GeTe. Further, interestingly, the phonon-isotope scattering rate ($\tau_{I}^{-1}(\omega)$) has been found to act as an indicator of phonon vacancy scattering rates ($\tau_{V}^{-1}$) with different vacancy concentration, for controlling the hydrodynamic regime for GeTe. These findings can help to demystify the unconventional hydrodynamic behavior in other chalcogenide alloys in the future. 

\begin{acknowledgments}
This project has received funding from the European Union’s Horizon 2020 research and innovation program under Grant Agreement No. 824957 (“BeforeHand:” Boosting Performance of Phase Change Devices by Hetero- and Nanostructure Material Design).
\end{acknowledgments}

\nocite{*}

\providecommand{\noopsort}[1]{}\providecommand{\singleletter}[1]{#1}%
\end{document}